\documentclass[prc,twocolumn,tightenlines
,amssymb,aps,nobibnotes,superscriptaddress,showpacs,balancelastpage,floatfix]{revtex4}
\usepackage{graphicx}
\begin{document}
\title{Consistent description of nuclear charge radii and electric monopole transitions}
\author{S.~Zerguine}
\affiliation{Department of Physics, PRIMALAB Laboratory, University of Batna,
Avenue Boukhelouf M El Hadi, 05000 Batna, Algeria}

\author{P.~Van~Isacker}
\affiliation{Grand Acc\'el\'erateur National d'Ions Lourds, CEA/DSM--CNRS/IN2P3,
BP~55027, F-14076 Caen Cedex 5, France}

\author{A.~Bouldjedri}
\affiliation{Department of Physics, PRIMALAB Laboratory, University of Batna,
Avenue Boukhelouf M El Hadi, 05000 Batna, Algeria}

\begin{abstract}
A systematic study of energy spectra
throughout the rare-earth region (even-even nuclei from $_{58}$Ce to $_{74}$W)
is carried out in the framework of the interacting boson model (IBM),
leading to an accurate description
of the spherical-to-deformed shape transition in the different isotopic chains.
The resulting IBM Hamiltonians are then used for the calculation
of nuclear charge radii (including isotope and isomer shifts) and electric monopole transitions
with consistent operators for the two observables.
The main conclusion of this study is that
an IBM description of charge radii and electric monopole transitions
is possible for most of the nuclei considered
but that it breaks down in the tungsten isotopes.
It is suggested that this failure
is related to hexadecapole deformation.
\end{abstract}

\maketitle

\section{Introduction}
\label{s_intro}
The structural properties of excited $0^+$ states in deformed even-even nuclei
have been the subject of a long controversy.
According to the geometric model of Bohr and Mottelson~\cite{Bohr75},
a nucleus with an elipsoidal equilibrium shape
may undergo oscillations of two different types, $\beta$ and $\gamma$.
The first type of vibration preserves axial symmetry
while the second allows excursions toward triaxial shapes.
These vibrations should be combined with rotations
exhibited by the deformed, vibrating nucleus
to yield a rotation-vibration spectrum.
The geometric model of Bohr and Mottelson, therefore,
predicts an axially deformed nucleus to display
a spectrum of rotational bands built on top of vibrational excitations.
The lowest in energy is the ground-state rotational band with $K^\pi=0^+$
({\it i.e.}, of which the projection of the angular momentum
on the axis of symmetry is zero)
which corresponds to no intrinsic excitation.
Next in energy are the $\beta$- and $\gamma$-vibrational bands
which correspond to one intrinsic excitation of the $\beta$ or $\gamma$ type,
characterized by a rotational band with $K^\pi=0^+$ or $2^+$, respectively.
At higher energies still, the geometric model predicts rotational bands 
built on multiple excitations of $\beta$ and/or $\gamma$ phonons.

While $\gamma$-vibrational bands
are an acknowledged feature of deformed nuclei,
such is not the case for $\beta$-vibrational bands.
Confusion arises because excited $0^+$ states in nuclei
can be of many different characters,
such as pairing isomers~\cite{Ragnarsson76},
two-quasi-particle excitations~\cite{Soloviev86},
or so-called intruder states
that arise through the mechanism of shape coexistence~\cite{Heyde11}.
A careful analysis of the observed properties of excited $0^+$ states
seems to indicate that very few indeed
satisfy all criteria proper to a $\beta$-vibrational state~\cite{Garrett01}.
In particular, although this observation is obfuscated by a lack of reliable data,
very few $0^+$ states decay to the ground state
by way of an electric monopole transition of sizable strength~\cite{Wood99},
as should be the case for a $\beta$ vibration~\cite{Reiner61}.
It is therefore not surprising
that alternative interpretations of excited $0^+$ states in deformed nuclei,
either as pairing isomers~(see, {\it e.g.}, Refs.~\cite{Kulp03,Kulp05})
or through shape coexistence and configuration mixing (see, {\it e.g.}, Ref.~\cite{Kulp08})
have gained advocates over recent years.

The purpose of this paper is to examine to what extent
a purely collective interpretation of nuclear $0^+$ levels
is capable of yielding a coherent and consistent description
of observed charge radii and electric monopole transitions~\cite{Zerguine08}.
As noted above, very few measured electric monopole transitions
satisfy the criteria proper to a matrix element from the ground-state to the $\beta$-vibrational band
and the present attempt therefore might seem doomed to failure.
However, collective excitations of nuclei
can also be described with the interacting boson model (IBM)
of Arima and Iachello~\cite{Arima76,Arima78,Arima79},
where they are modeled
in terms of a constant number of $s$ and $d$ (and sometimes $g$) bosons
which can be thought of as correlated pairs of nucleons
occupying valence shell-model orbits
coupled to angular momentum $\ell=0$ and 2 (and 4), respectively.
One of the advantages of the IBM
is that a connection with the shell model~\cite{Otsuka78}
as well as with the geometric model~\cite{Dieperink80,Ginocchio80a,Bohr80}
has been established.
In particular, one of its dynamical symmetries, the SU(3) limit~\cite{Arima78},
displays energies reminiscent of the rotation-vibration spectrum
of the geometric model.
It has also been shown, however,
that the first-excited $0^+$ state in the SU(3) limit of the IBM
has not exactly a $\beta$-vibrational character
but is a complicated mixture of intrinsic $\beta$ and $\gamma$ vibrations~\cite{Casten88}.
The main purpose of this paper is to show
that a collective interpretation of excited $0^+$ states with the IBM
is not inconsistent with the electric monopole data,
as observed in the rare-earth region.

The outline of this paper as follows.
In Sect.~\ref{s_r2e0} the ground is prepared
by discussing charge radii and electric monopole transitions
in the context of different models
that are applied to even-even nuclei in the rare-earth region
from Ce to W in Sect.~\ref{s_application}.
A qualitative explanation of the failure of this approach in the W isotopes
is offered in Sect.~\ref{s_g} by invoking effects of hexadecapole deformation.
Finally, Sect.~\ref{s_conc} summarizes the conclusions of this work.

\section{Charge radii and electric monopole transitions}
\label{s_r2e0}
Electric monopole (E0) transitions between nuclear levels
proceed mainly by internal conversion
with no transfer of angular momentum to the ejected electron.
If the energy of the transition is greater than $2m_{\rm e}c^2$
(where $m_{\rm e}$ is the mass of the electron),
they can occur via electron-positron pair creation.
A less probable de-excitation mode
which can proceed via an E0 transition is two-photon emission.
It is not {\it a priori} clear why a connection
exists between charge radii and E0 transitions.
In fact, the argument is rather convoluted
and we begin this section by recalling it.
The argument can be generalized to effective operators,
leading to a relation between charge radii and E0 transitions
which forms the basis of the present study.

\subsection{Relation between effective operators
for charge radii and electric monopole transitions}
\label{ss_relation}
The total probability for an E0 transition between initial and final states
$|{\rm i}\rangle$ and $|{\rm f}\rangle$
can be written as the product of an electronic factor $\Omega$
and a nuclear factor $\rho({\rm E0})$,
the latter being equal to~\cite{Church56}
\begin{equation}
\rho({\rm E0})=
\langle{\rm f}|
\sum_{k\in\pi}\left[\left(\frac{r_k}{R}\right)^2-
\sigma\left(\frac{r_k}{R}\right)^4+\cdots\right]
|{\rm i}\rangle,
\label{e_rhoi}
\end{equation}
with $R=r_0A^{1/3}$
and where the summation runs over the $Z$ protons ($\pi$) in the nucleus.
The coefficient $\sigma$ depends on the assumed nuclear charge distribution
but in any reasonable case it is smaller than 0.1.
The second term in Eq.~(\ref{e_rhoi}) therefore can be neglected
if the leading term is not too small~\cite{Church56}.
In this approximation we have
\begin{equation}
\rho({\rm E0})\approx
\frac{1}{R^2}
\langle{\rm f}|
\sum_{k\in\pi}r_k^2
|{\rm i}\rangle.
\label{e_rhoa}
\end{equation}

On the other hand,
the mean-square charge radius of a state $|{\rm s}\rangle$
is given by
\begin{equation}
\langle r^2 \rangle_{\rm s}=
\frac{1}{Z}\langle{\rm s}|\sum_{k\in\pi}r^2_k|{\rm s}\rangle.
\label{e_r2a}
\end{equation}
This is an appropriate expression
insofar that a realistic $A$-body wave function
is used for the state $|{\rm s}\rangle$.
For the heavy nuclei considered here
the construction of such realistic wave function is difficult
and recourse to an effective charge radius operator $\hat T(r^2)$ should be taken. 
In particular, if neutrons are assigned an effective charge,
the polarization of the protons due to the neutrons is `effectively' taken into account,
giving rise to changes in the charge radius $\langle r^2\rangle$ with neutron number. 
The generalization of the  expression~(\ref{e_r2a})
can therefore be written as
\begin{eqnarray}
\langle r^2 \rangle_{\rm s}&\equiv&
\langle{\rm s}|\hat T(r^2)|{\rm s}\rangle=
\frac{1}{e_{\rm n}N+e_{\rm p}Z}
\langle{\rm s}|\sum_{k=1}^Ae_kr^2_k|{\rm s}\rangle
\nonumber\\&=&
\frac{1}{e_{\rm n}N+e_{\rm p}Z}
\langle{\rm s}|
e_{\rm n}\sum_{k\in\nu}r^2_k+
e_{\rm p}\sum_{k\in\pi}r^2_k
|{\rm s}\rangle,
\label{e_r2}
\end{eqnarray}
where the first summation runs over all $A$ nucleons
while the second and third summations run
over neutrons ($\nu$) and protons ($\pi$) only,
and where $e_{\rm n}$ ($e_{\rm p}$) is the effective neutron (proton) charge.
If bare nucleon charges are taken ($e_{\rm n}=0$  and $e_{\rm p}=e$),
the summation extends over protons only
and the original expression~(\ref{e_r2a}) is recovered.
If equal nucleon charges are taken ($e_{\rm n}=e_{\rm p}$),
Eq.~(\ref{e_r2}) is appropriate for the matter radius.

In the approximation $\sigma\approx0$,
the starting expressions~(\ref{e_rhoa}) and~(\ref{e_r2a})
for the nuclear E0 transition strength and the mean-square charge radius
are identical (up to the constants $R^2$ and $Z$).
It is therefore natural to follow the same argument
as used for the charge radius
for the construction of a generalized E0 transition operator,
leading to the expression~\cite{Kantele84}
\begin{equation}
\hat T({\rm E0})=
\sum_{k=1}^Ae_kr^2_k=
e_{\rm n}\sum_{k\in\nu}r^2_k+e_{\rm p}\sum_{k\in\pi}r^2_k.
\label{e_e0}
\end{equation}
In terms of this operator,
the dimensionless quantity $\rho({\rm E0})$, defined in Eq.~(\ref{e_rhoa})
and referred to as the monopole strength,
is given by
\begin{equation}
\rho({\rm E0})=\frac{\langle{\rm f}|\hat T({\rm E0})|{\rm i}\rangle}{eR^2}.
\label{e_rho}
\end{equation}
Since the matrix element~(\ref{e_rho}) is known up to a sign only,
usually $\rho^2({\rm E0})$ is quoted.

The basic hypothesis of the present study
is to assume that {\em the effective nucleon charges
in the charge radius and E0 transition operators are the same}.
If this is so, comparison of Eqs.~(\ref{e_r2}) and~(\ref{e_e0}) leads to the relation
\begin{equation}
\hat T({\rm E0})=(e_{\rm n}N+e_{\rm p}Z)\hat T(r^2).
\label{e_e0r2}
\end{equation}
This is a general relation between the effective operators
used for the calculation of charge radii and E0 transitions,
which is applied throughout this study.

\subsection{Charge radii and electric monopole transitions
in the interacting boson model}
\label{ss_ibm}
Equation~(\ref{e_e0r2}) can, in principle, be tested in the framework of any model. 
This endeavor is difficult in the context of the nuclear shell model
because realistic wave functions,
appropriate for the calculation of $\langle r^2\rangle$ or E0 matrix elements,
are hard to come by for heavy nuclei.
As an alternative we propose here
to test the implied correlation with the use of a simpler model,
namely the IBM~\cite{Arima76,Arima78,Arima79}.
This requires that all states involved
[{\it i.e.}, $|{\rm s}\rangle$  in Eq.~(\ref{e_r2}),
and $|{\rm i}\rangle$ and $|{\rm f}\rangle$ in Eq.~(\ref{e_rho})]
are collective in character
and can be described by the IBM.

In the \mbox{IBM-1},
where no distinction is made between neutron and proton bosons,
the charge radius operator
is taken as the most general scalar expression
that is linear in the generators of U(6)~\cite{Iachello87},
\begin{equation}
\hat T(r^2)=
\langle r^2\rangle_{\rm c}+
\alpha N_{\rm b}+
\eta\frac{\hat n_d}{N_{\rm b}},
\label{e_r2sd}
\end{equation}
where $N_{\rm b}$ is the total boson number
(the customary notation $N$ is not used here
to avoid confusion with the neutron number),
$\hat n_d$ is the $d$-boson number operator,
and $\alpha$, $\eta$ are parameters with units of length$^2$.
The first term in Eq.~(\ref{e_r2sd}), $\langle r^2\rangle_{\rm c}$,
is the charge radius of the core nucleus.
The second term accounts for the (locally linear) increase
in the charge radius due to the addition of two nucleons
({\it i.e.}, neutrons since isotope shifts are considered in this study).
The boson number $N_{\rm b}$
is the number of pairs of valence particles or holes (whichever is smaller)
counted from the nearest closed shells for neutrons and protons.
If the bosons are particle-like,
the addition of two nucleons
corresponds to an {\em increase} of $N_{\rm b}$ by one
and $\alpha$ is positive.
In contrast, if the bosons are hole-like,
the addition of two nucleons
corresponds to a {\em decrease} of $N_{\rm b}$ by one
and $\alpha$ is negative.
Therefore, care should be taken
to change the sign of $\alpha$ at mid-shell~\cite{Iachello87}.
The third term in Eq.~(\ref{e_r2sd})
stands for the contribution to the charge radius due to deformation.
It is identical to the one given in Ref.~\cite{Iachello87}
but for the factor $1/N_{\rm b}$.
This factor is included here
because it is the {\em fraction} $\langle\hat n_d\rangle/N_{\rm b}$
which is a measure of the quadrupole deformation
($\beta_2^2$ in the geometric collective model)
rather than the matrix element $\langle\hat n_d\rangle$ itself.
Since the inclusion of $1/N_{\rm b}$ is non-standard,
also results without this factor will be given in the following,
that is, with the charge radius operator
\begin{equation}
\hat T'(r^2)=
\langle r^2\rangle_{\rm c}+
\alpha'N_{\rm b}+
\eta'\hat n_d.
\label{e_r2sda}
\end{equation}

Once the form of the charge radius operator is determined,
that of the E0 transition operator follows from Eq.~(\ref{e_e0r2}).
In the \mbox{IBM-1} the E0 transition operators are therefore
\begin{equation}
\hat T({\rm E0})=
(e_{\rm n}N+e_{\rm p}Z)
\eta\frac{\hat n_d}{N_{\rm b}},
\label{e_e0sd}
\end{equation}
or
\begin{equation}
\hat T'({\rm E0})=
(e_{\rm n}N+e_{\rm p}Z)\eta'\hat n_d.
\label{e_e0sda}
\end{equation}
Since for E0 transitions the initial and final states are different,
neither the constant $\langle r^2\rangle_{\rm c}$
nor $\alpha N_{\rm b}$ in Eq.~(\ref{e_r2sd})
or $\alpha'N_{\rm b}$ in Eq.~(\ref{e_r2sda}) contribute to the transition,
so they can be omitted from the E0 operators.

Two other quantities can be derived from charge radii,
namely isotope and isomer shifts.
The former measures the difference in charge radius of neighboring isotopes.
For the difference between even-even isotopes
one finds from Eq.~(\ref{e_r2sd}) 
\begin{eqnarray}
\Delta \langle r^2\rangle^{(A)}&\equiv&
\langle r^2\rangle_{0_1^+}^{(A+2)}-\langle r^2\rangle_{0_1^+}^{(A)}
\nonumber\\&=&
|\alpha|+\eta
\left(\langle\frac{\hat n_d}{N_{\rm b}}\rangle_{0_1^+}^{(A+2)}-
\langle\frac{\hat n_d}{N_{\rm b}}\rangle_{0_1^+}^{(A)}\right).
\label{e_ips}
\end{eqnarray}
The occurrence of the absolute value $|\alpha|$
is due to the interpretation of the bosons
as pairs of particles or holes, as discussed above.
Isomer shifts are a measure of the difference in charge radius
between an excited (here the $2_1^+$) state
and the ground state, and are given by
\begin{eqnarray}
\delta\langle r^2\rangle^{(A)}&\equiv&
\langle r^2\rangle^{(A)}_{2_1^+}-\langle r^2\rangle^{(A)}_{0_1^+}
\nonumber\\&=&
\eta
\left(\langle\frac{\hat n_d}{N_{\rm b}}\rangle_{2_1^+}^{(A)}-
\langle\frac{\hat n_d}{N_{\rm b}}\rangle_{0_1^+}^{(A)}\right).
\label{e_ims}
\end{eqnarray}
Similar formulas hold for the charge radius operator~(\ref{e_r2sda})
in terms of the parameters $\alpha'$ and $\eta'$.
For ease of notation,
the superscript $(A)$ in the isotope and isomer shifts
shall be suppressed in the following.

\subsection{Estimate of the coefficients $\alpha$ and $\eta$}
\label{ss_estimate1}
Although the coefficients $\alpha$ and $\eta$ in Eq.~(\ref{e_r2sd})
will be treated as parameters
and fitted to data on charge radii and E0 transitions,
it is important to have an estimate of their order of magnitude.
The term in $\alpha$ increases with particle number
and therefore can be associated with the `standard' isotope shift.
This standard contribution to the charge radius
is given by~\cite{Bohr69}
\begin{equation}
\langle r^2\rangle^{(A)}_{0_1^+,{\rm std}}\approx
\frac{3}{5}r_0^2A^{2/3}.
\label{e_r2std}
\end{equation}
The term in $\eta$ stands for the contribution to the nuclear radius due to deformation.
For a quadrupole deformation it is estimated to be~\cite{Bohr69}
\begin{equation}
\langle r^2\rangle^{(A)}_{0_1^+,{\rm def}}\approx
\frac{5}{4\pi}\beta_2^2\langle r^2\rangle^{(A)}_{0_1^+,{\rm std}}\approx
\frac{3}{4\pi}\beta_2^2r_0^2A^{2/3},
\label{e_r2def}
\end{equation}
where $\beta_2$ is the quadrupole deformation parameter of the geometric model.

The estimate of $|\alpha|$ follows from
\begin{eqnarray}
|\alpha|&=&
\Delta\langle r^2\rangle^{(A)}_{\rm std}\approx
\frac{3}{5}r_0^2\left((A+2)^{2/3}-A^{2/3}\right)
\nonumber\\&\approx&
\frac{4}{5}r_0^2A^{-1/3},
\label{e_esta}
\end{eqnarray}
which for the nuclei considered here ($A\sim150$) gives $|\alpha|\approx0.2$~fm$^2$.
The estimate of $\eta$ can be obtained
by associating the deformation contribution~(\ref{e_r2def})
with the expectation value of $\langle\hat n_d\rangle_{0^+_1}$ in the IBM.
This leads to the relation
\begin{equation}
\eta\frac{{\bar\beta_2}^2}{1+{\bar\beta_2}^2}\approx
\frac{4}{3}{\bar\beta_2}^2r_0^2N_{\rm b}^2A^{-4/3},
\label{e_estb}
\end{equation}
where use has been made of the approximate correspondence 
$\beta_2\approx(4N_{\rm b}/3A)\sqrt{\pi}\bar\beta_2$
between the quadrupole deformations $\beta_2$ and $\bar\beta_2$
in the geometric model and in the IBM, respectively~\cite{Ginocchio80b}.
The relation~(\ref{e_estb}) yields the estimate
\begin{equation}
\eta\approx
\frac{4}{3}(1+{\bar\beta_2}^2)r_0^2N_{\rm b}^2A^{-4/3}.
\label{e_estb2}
\end{equation}
For typical values of $N_{\rm b}\sim10$ and $A\sim150$
this gives a range of possible $\eta$ values between 0.25 and 0.75~fm$^2$,
corresponding to weakly deformed ($\bar\beta_2\ll1$)
and strongly deformed ($\bar\beta_2\approx\sqrt2$) nuclei, respectively.

Similar estimates can be derived for the coefficients
in the alternative form~(\ref{e_r2sda}) of the charge radius operator.
The estimate for $|\alpha'|$ is identical to Eq.~(\ref{e_esta})
while the one for $\eta'$ differs by a factor $N_{\rm b}$.

\subsection{Estimate of the effective charges}
\label{ss_estimate2}
The consistent definition of operators for charge radii and E0 transitions
leads to the introduction of neutron and proton effective charges for {\em both} operators,
as opposed to previous treatments
where this was done for E0 transitions only.
This opens the possibility to obtain an estimate
of the effective charges $e_{\rm n}$ and $e_{\rm p}$
from data for charge radii which are widely available.

A simple estimate can be obtained in a Hartree-Fock approximation
with harmonic-oscillator single-particle wave functions.
For the Hartree-Fock ground state $|{\rm gs}\rangle$ of a nucleus,
simple counting arguments of the degeneracies
of the three-dimensional harmonic oscillator
(see Sect.~2.2 of Ref.~\cite{Brussaard77}),
lead to the following expression for the expectation value
of the neutron part of the charge radius operator:
\begin{equation}
\langle{\rm gs}|\sum_{k\in\nu}r^2_k|{\rm gs}\rangle=
\frac{3^{4/3}}{4}N^{4/3}b^2=
\frac{3\cdot2^{1/3}}{5}r_0^2N^{4/3}A^{1/3},
\label{e_estr2n}
\end{equation}
where $b$ is the oscillator length of the harmonic oscillator
for which the estimate $b=2^{7/6}3^{-1/6}5^{-1/2}r_0A^{1/6}$
is used~\cite{Brussaard77}.
Equation~(\ref{e_estr2n})
and its equivalent for the proton part of the charge radius operator,
can then be combined to yield
\begin{equation}
\langle r^2\rangle_{\rm gs}=
\frac{3\cdot2^{1/3}}{5}r_0^2
\frac{(e_{\rm n}N^{4/3}+e_{\rm p}Z^{4/3})A^{1/3}}{e_{\rm n}N+e_{\rm p}Z}.
\label{e_estr2}
\end{equation}
The separate effective charges $e_{\rm n}$ and $e_{\rm p}$
cannot be obtained from a fit of this expression
to the data on charge radii,
but the ratio $e_{\rm n}/e_{\rm p}$ can.
There exists, unfortunately, a strong dependence of this ratio
on the value used for $r_0$,
as will be discussed in Subsect.~\ref{ss_e0}.

\subsection{Relating charge radii and electric monopole transitions
through configuration mixing}
\label{ss_coexistence}
The idea of correlating observed E0 transitions with charge radii 
is not new and, in particular, was already proposed by Wood {\it et al.}~\cite{Wood99}.
These authors tested this proposal in the context of a model
assuming mixing between two different coexisting configurations.
Since in Sect.~\ref{s_application} results are quoted for the few nuclei
where this model has been applied,
a brief reminder of the method is given in this subsection.

Assume that the initial and final levels in the E0 transition
are orthogonal mixtures of some states $|\alpha_1J\rangle$ and $|\alpha_2J\rangle$,
so that
\begin{eqnarray}
|{\rm i}J\rangle&=&a_J|\alpha_1J\rangle+b_J|\alpha_2J\rangle,
\nonumber\\
|{\rm f}J\rangle&=&a_J|\alpha_2J\rangle-b_J|\alpha_1J\rangle.
\label{e_mix}
\end{eqnarray}
The labels $\alpha_1$ and $\alpha_2$
refer to a different intrinsic structure
for the two sets of states (typically two bands),
the members of which are additionally characterized
by their angular momentum $J$.
The $a_J$ and $b_J$ are mixing coefficients
that satisfy $a_J^2+b_J^2=1$.
If one assumes that no E0 transition is allowed
between states with a different intrinsic structure,
$\langle\alpha_1J|\hat T({\rm E0})|\alpha_2J\rangle\approx0$,
it follows that
\begin{eqnarray}
\lefteqn{\langle{\rm f}J|\hat T({\rm E0})|{\rm i}J\rangle}
\nonumber\\&\approx&
a_Jb_J\left(
\langle\alpha_2J|\hat T({\rm E0})|\alpha_2J\rangle-
\langle\alpha_1J|\hat T({\rm E0})|\alpha_1J\rangle\right)
\nonumber\\&=&
a_Jb_J
(e_{\rm n}N+e_{\rm p}Z)\left(
\langle r^2\rangle_{\alpha_2J}-
\langle r^2\rangle_{\alpha_1J}\right),
\label{e_mixint}
\end{eqnarray}
where the last equality is due to Eq.~(\ref{e_e0r2}).
Furthermore, it must be hoped that
the difference in $\langle r^2\rangle$ appearing in Eq.~(\ref{e_mixint})
can be identified with a measured isotope shift $\Delta\langle r^2\rangle$
between nuclei somewhere in the neighborhood
and, therefore, that the ground states of two nuclei in the neighborhood
can be identified with the unmixed intrinsic structures $\alpha_1$ and $\alpha_2$.
Finally, it must be assumed
that this difference in $\langle r^2\rangle$
does not depend significantly on $J$
or, equivalently, that isomer shifts are identical
for the intrinsic structures $\alpha_1$ and $\alpha_2$.
With this first set of assumptions
the following relation holds:
\begin{equation}
\rho^2_J({\rm E0})=
a_J^2b_J^2\frac{(e_{\rm n}N+e_{\rm p}Z)^2}{e^2R^4}
[\Delta\langle r^2\rangle]^2,
\label{e_mixrho}
\end{equation}
which reduces to the result of Wood {\it et al.}~\cite{Wood99}
if bare nucleon charges are taken.
According to this equation
the $J$ dependence of the E0 strength
is contained in the coefficients $a_J$ and $b_J$,
which can be obtained from a two-state mixing calculation.
An additional input into the calculation, therefore,
is the relative position in energy
of the intrinsic structures $\alpha_1$ and $\alpha_2$
and the size of the mixing matrix element
(itself assumed to be independent of $J$).
These are the essential ingredients of the calculations
reported by Kulp {\it et al.}~\cite{Kulp08}
of which the results are quoted below.

\section{Systematic study of nuclei in the rare-earth region}
\label{s_application}
To test the relation between charge radii and E0 transitions,
proposed in the previous section, 
a systematic study of all even-even isotopic chains
from Ce to W is carried out.
This analysis requires the knowledge of structural information
concerning the ground states and excited states
which here is obtained by adjusting an \mbox{IBM-1} Hamiltonian
to observed spectra in the rare-earth region.

\subsection{Hamiltonian and energy spectra}
\label{ss_hamilt}
All isotope series in the rare-earth region from $Z=58$ to $Z=74$
have the particularity to vary from spherical to deformed shapes 
and to display systematically a shape phase transition.
Such nuclear behavior can be parametrized
in terms of a simplified \mbox{IBM-1} Hamiltonian
which can be represented on the so-called Casten triangle~\cite{Casten81},
as demonstrated for rare-earth nuclei by Mc~Cutchan {\it et al.}~\cite{Cutchan04}.
Alternatively, as shown by Garc\'\i a-Ramos {\it et al.}~\cite{Garcia03},
the same region of the nuclear chart
can be described with the full IBM Hamiltonian.
The latter approach is adopted here
and a general one- and two-body Hamiltonian is considered
which is written in multipole form as~\cite{Iachello87}
\begin{equation}
\hat H=
\epsilon_d\hat n_d+
a_0\hat P_+\cdot\hat P_-+
a_1\hat L\cdot\hat L+
a_2\hat Q\cdot\hat Q+
a_3\hat T_3\cdot\hat T_3+
a_4\hat T_4\cdot\hat T_4,
\label{e_hgen} 
\end{equation}
where $\hat n_d$ is the $d$-boson number operator,
$\hat P_+$ ($\hat P_-$) is a boson-pair creation (annihilation) operator,
$\hat L$ is the angular momentum operator,
and $\hat Q$, $\hat T_3$, and $\hat T_4$
are quadrupole, octupole, and hexadecapole operators, respectively.
(For the definitions of these operators in terms of $s$ and $d$ bosons,
see Sect.~1.4.7 of Iachello and Arima~\cite{Iachello87}.)
If only excitation and no absolute energies are considered,
the expression~(\ref{e_hgen}) defines the most general Hamiltonian
with one- and two-body interactions between the bosons
in terms of six parameters.
The parameter $\chi$ in the quadrupole operator $\hat Q$
can be arbitrarily chosen
(as long as it is different from zero)
and is fixed here to $-\sqrt{7}/2$ for all nuclei.

Although reasonable results are obtained
with constant parameters for an entire chain of isotopes,
the spherical-to-deformed transition is better described
if at least one parameter is allowed to vary with boson number $N_{\rm b}$.
Garc\'\i a-Ramos {\it et al.}~\cite{Garcia03} followed the procedure
to vary the $d$-boson energy $\epsilon_d$ with $N_{\rm b}$.
A different method is followed here
by allowing the variation of the coefficient $a_2$
associated with the quadrupole term.
To reduce the number of parameters in the fit,
we assume that $a_2$ depends linearly on the quantity $N_\nu N_\pi/(N_\nu+N_\pi)$
where $N_\nu$ ($N_\pi$) is half the number of valence neutrons (protons)
particles or holes, whichever is smaller.
The argument for introducing such a dependence
is related to the importance of the neutron-proton interaction
as the deformation-driving mechanism in nuclei~\cite{Casten85}.
The parameter $a_2$ can then be decomposed into two terms as follows:
\begin{equation}
a_2=a'_2+\frac{N_\nu N_ \pi}{N_\nu+N_\pi}a''_2.
\label{e_qpar}
\end{equation}

\begin{table*}
\caption{Parameters in the Hamiltonian~(\ref{e_hgen})
and the rms deviation $\sigma$, in units of keV.}
\label{t_parh}
\begin{tabular}{@{}cc|crcrcrcrcrcrcrc|cr}
\hline
\hline
Isotopes&~~&&$\epsilon_d$&~~&$a_0$&~~&$a_1$&~~&$a'_2$&~~&$a''_2$&~~&$a_3$&~~&$a_4$&~~&&$\sigma$\\
\hline
$^{144-152}_{\phantom{000-0}58}$Ce&~&&
$1516.9$&~&$67.7$&~&$-8.6$&~&$-26.5$&~&$-2.2$&~&$-185.9$&~&$-113.1$&~&&$81$\\[0.5ex]
$^{146-156}_{\phantom{000-0}60}$Nd&~&&
$1701.0$&~&$55.9$&~&$-16.2$&~&$-17.2$&~&$-0.6$&~&$-78.1$&~&$-221.5$&~&&$124$\\[0.5ex]
$^{148-160}_{\phantom{000-0}62}$Sm&~&&
$944.2$&~&$-73.7$&~&$5.7$&~&$-0.2$&~&$-15.2$&~&$-227.4$&~&$70.0$&~&&$107$\\[0.5ex]
$^{150-162}_{\phantom{000-0}64}$Gd&~&&
$1857.2$&~&$69.1$&~&$-15.3$&~&$-8.4$&~&$-1.7$&~&$-52.8$&~&$-228.0$&~&&$115$\\[0.5ex]
$^{152-164}_{\phantom{000-0}66}$Dy&~&&
$1887.2$&~&$75.6$&~&$-12.3$&~&$-8.8$&~&$-0.7$&~&$-53.5$&~&$-219.2$&~&&$97$\\[0.5ex]
$^{154-170}_{\phantom{000-0}68}$Er&~&&
$1772.6$&~&$105.9$&~&$-10.9$&~&$-6.7$&~&$-0.9$&~&$-43.5$&~&$-222.0$&~&&$87$\\[0.5ex]
$^{156-176}_{\phantom{000-0}70}$Yb&~&&
$780.4$&~&$43.5$&~&$0.5$&~&$6.7$&~&$-5.2$&~&$-20.9$&~&$-41.4$&~&&$91$\\[0.5ex]
$^{160-182}_{\phantom{000-0}72}$Hf&~&&
$1061.7$&~&$62.5$&~&$-7.0$&~&$-7.0$&~&$-0.2$&~&$36.0$&~&$-128.2$&~&&$103$\\[0.5ex]
$^{164-190}_{\phantom{000-0}74}$W&~&&
$1068.4$&~&$73.0$&~&$-3.8$&~&$-7.0$&~&$-0.5$&~&$5.7$&~&$-136.7$&~&&$103$\\
\hline
\hline
\end{tabular}
\end{table*}
The parameters in the Hamiltonian~(\ref{e_hgen})
are determined from a least-squares fit
to levels of the ground-state band with $K^\pi=0^+$
and those of two more bands with $K^\pi=0^+$ and $K^\pi=2^+$
(`quasi-$\beta$' and `quasi-$\gamma$' bands).
Since two-quasi-particle excitations do not belong to the model space of the \mbox{IBM-1},
states beyond the backbend cannot be described in this version of the model
and for this reason only levels up to $J^\pi=10^+$ are included in the fit.
Similarly, near closed shells, excitations might be of single-particle character
and, therefore, nuclei with $N\leq 84$ are excluded from the energy fit.
Nevertheless, the Hamiltonian~(\ref{e_hgen}) with the parametrization~(\ref{e_qpar})
allows the extrapolation toward the $N=84$ isotopes,
which is needed for some of the isotope shifts calculated in the following.
The total number of nuclei included in the fit is 78
and the total number of excited levels is 846.
The parameters are summarized in Table~\ref{t_parh}
as well as the root-mean-square (rms) deviation $\sigma$ for each isotope chain
which typically is of the order of 100~keV.

\begin{figure*}
\centering
\includegraphics[width=5.8cm]{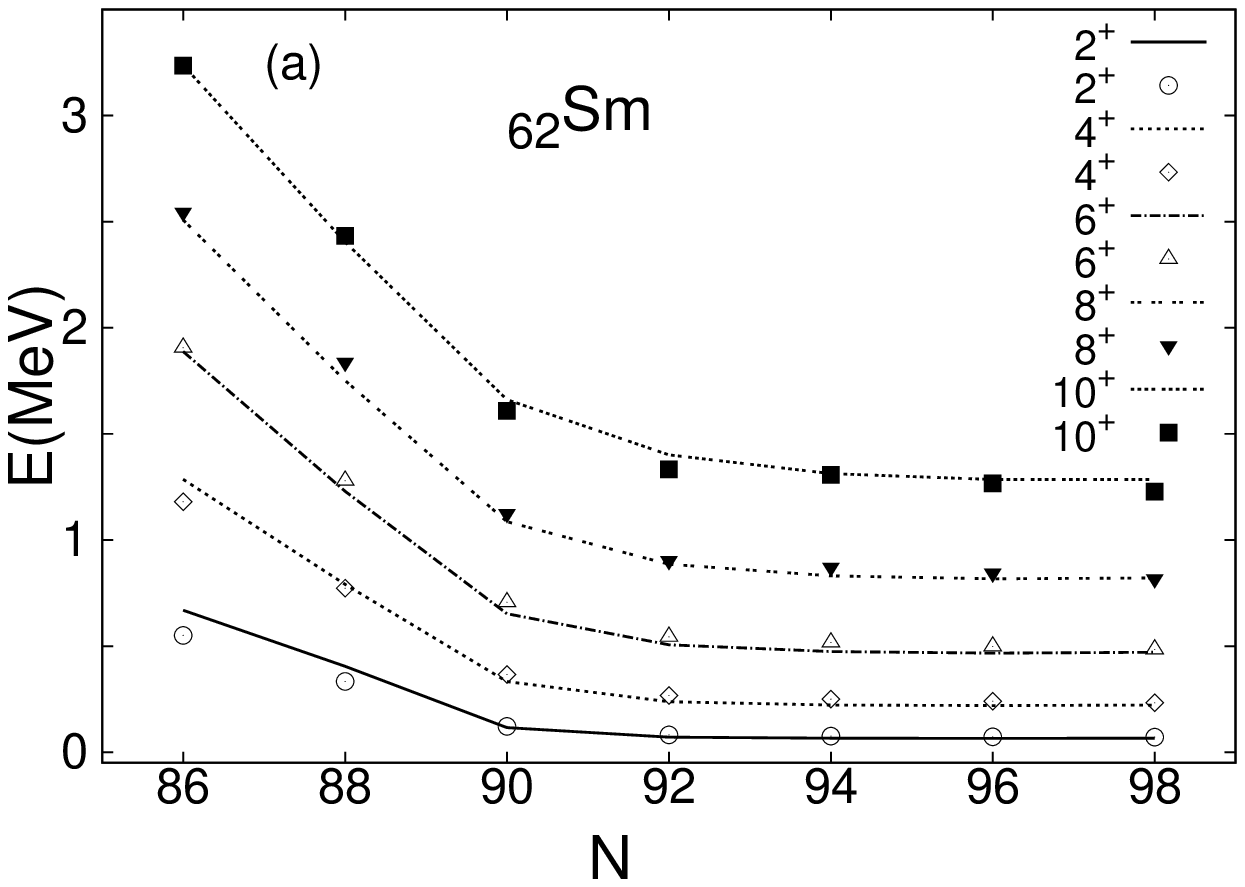}
\includegraphics[width=5.8cm]{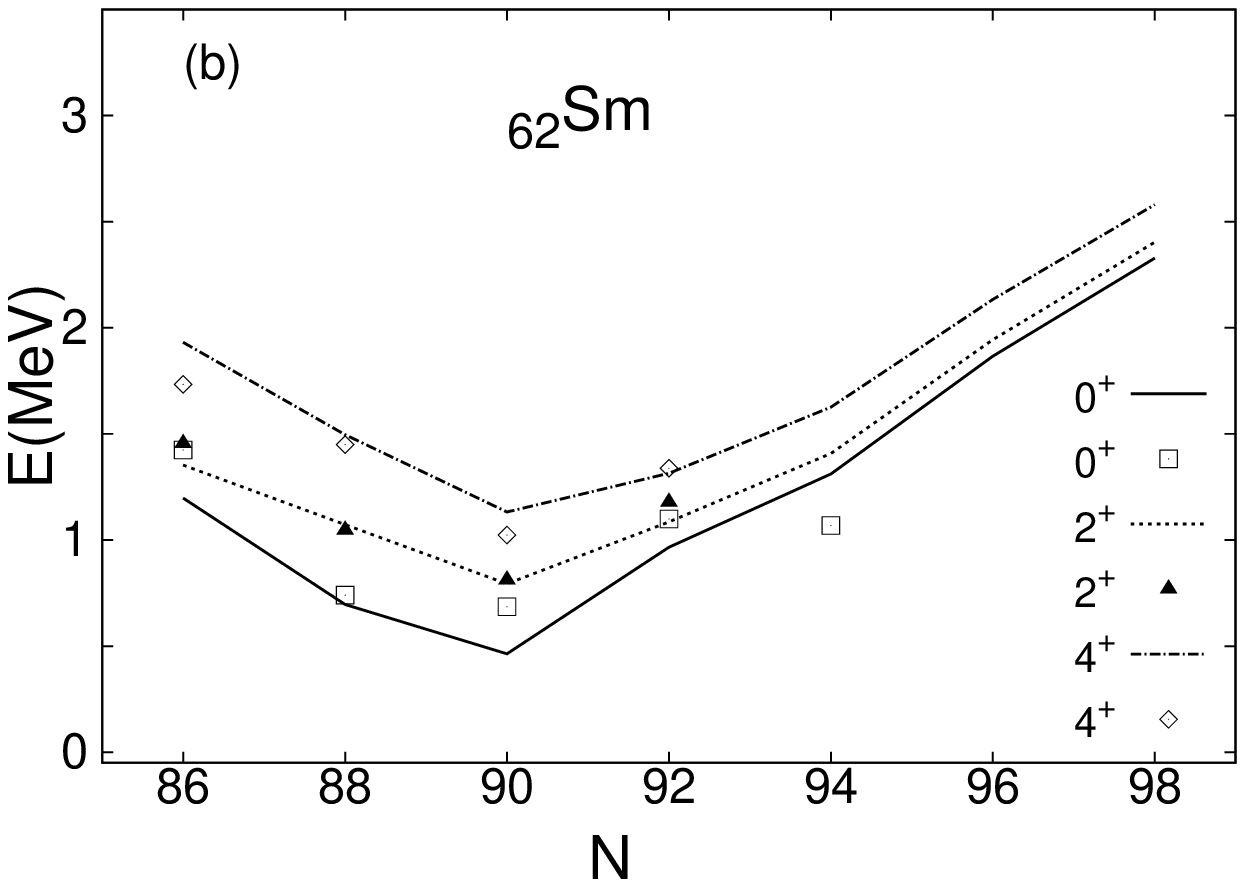}
\includegraphics[width=5.8cm]{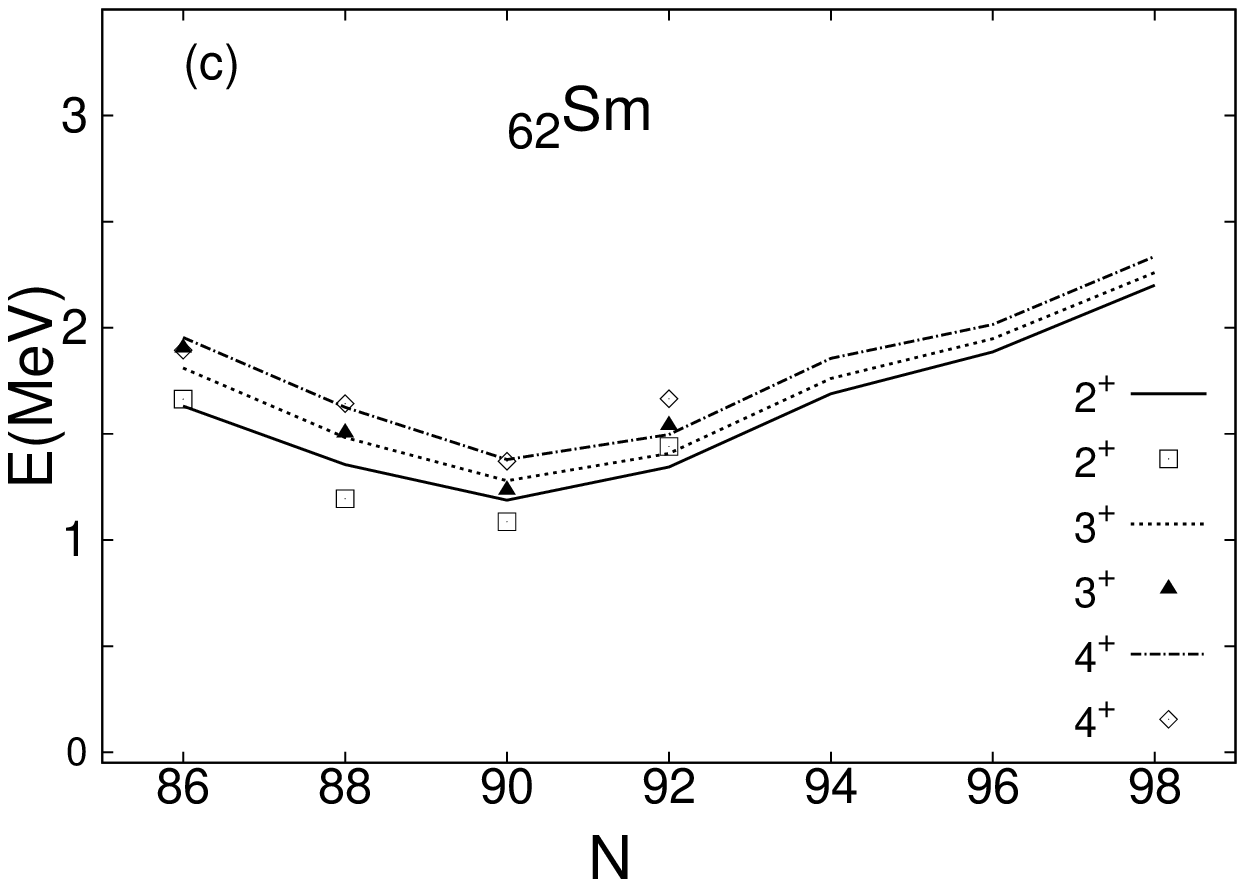}
\includegraphics[width=5.8cm]{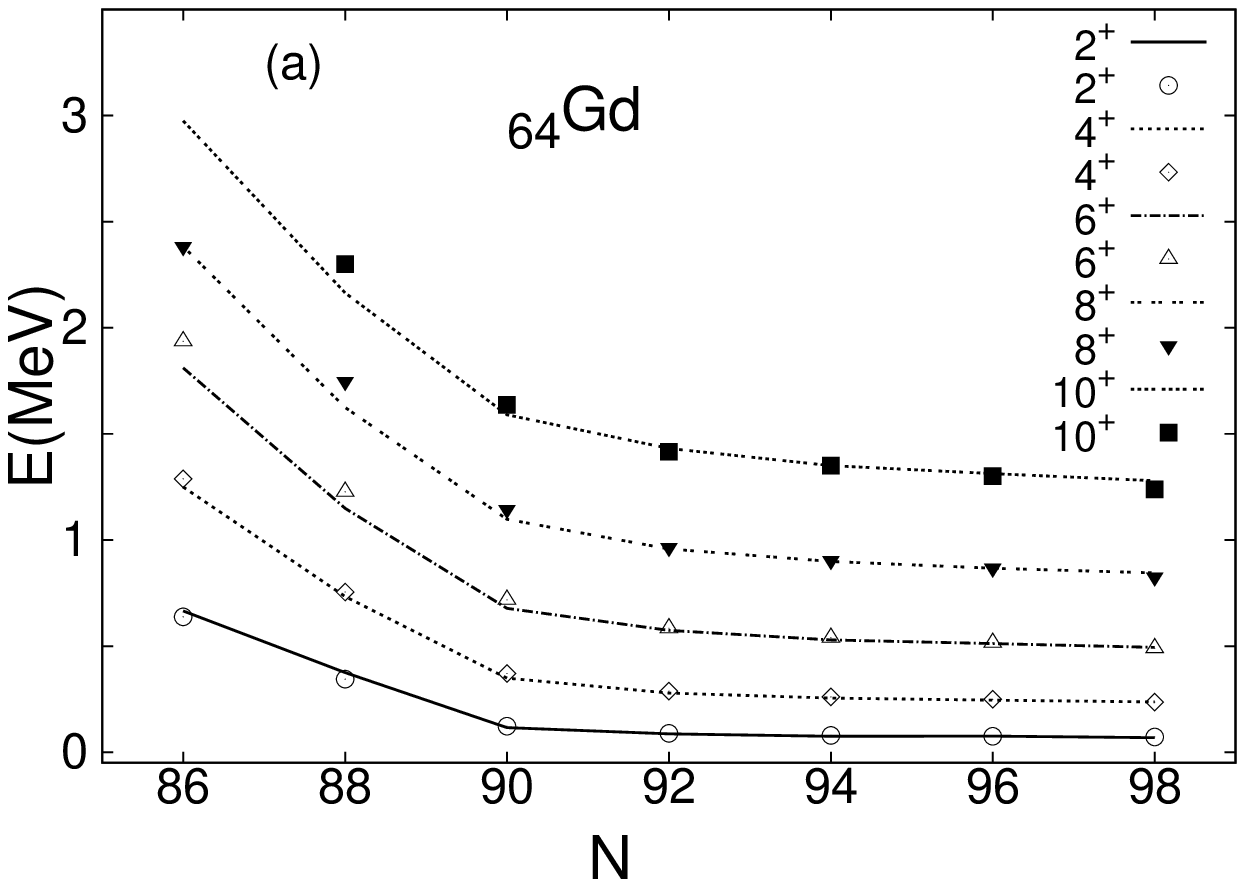}
\includegraphics[width=5.8cm]{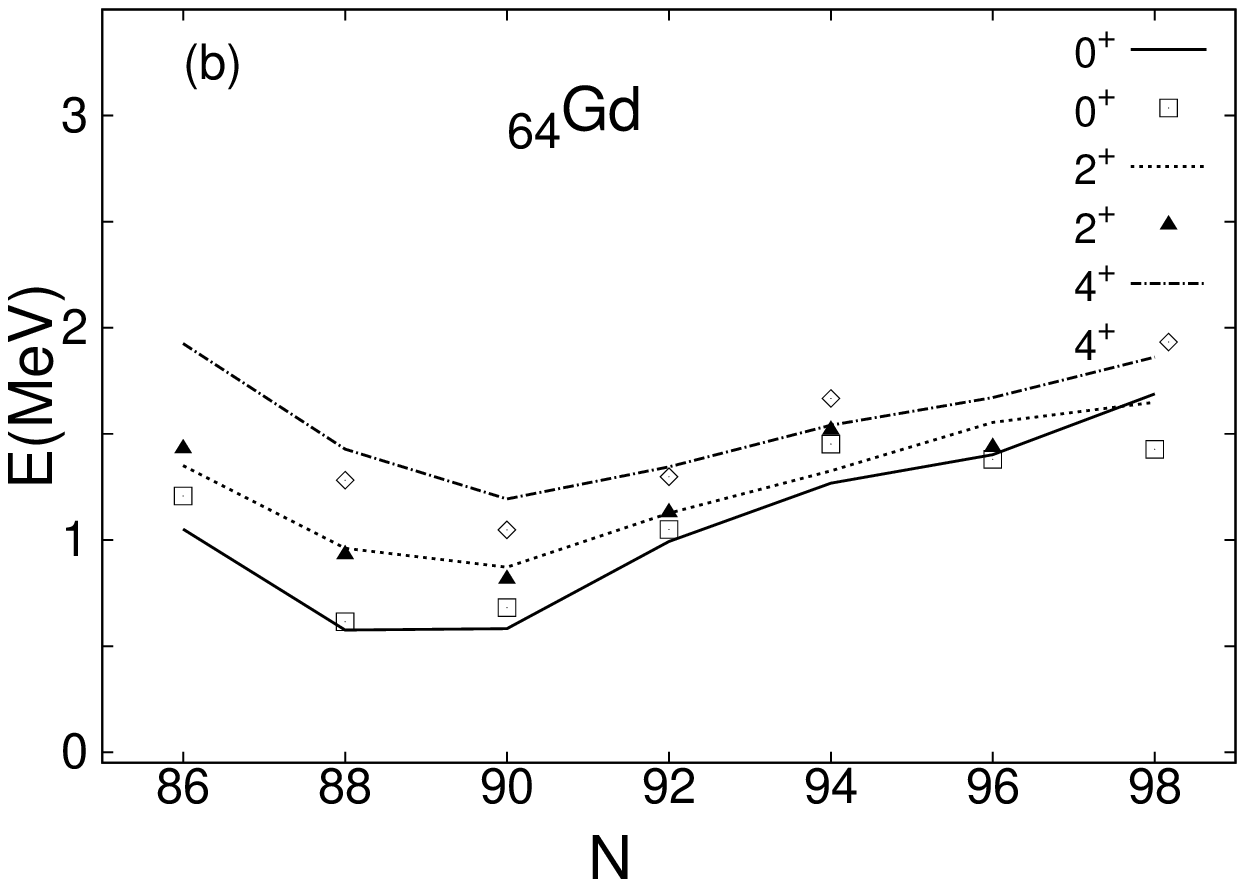}
\includegraphics[width=5.8cm]{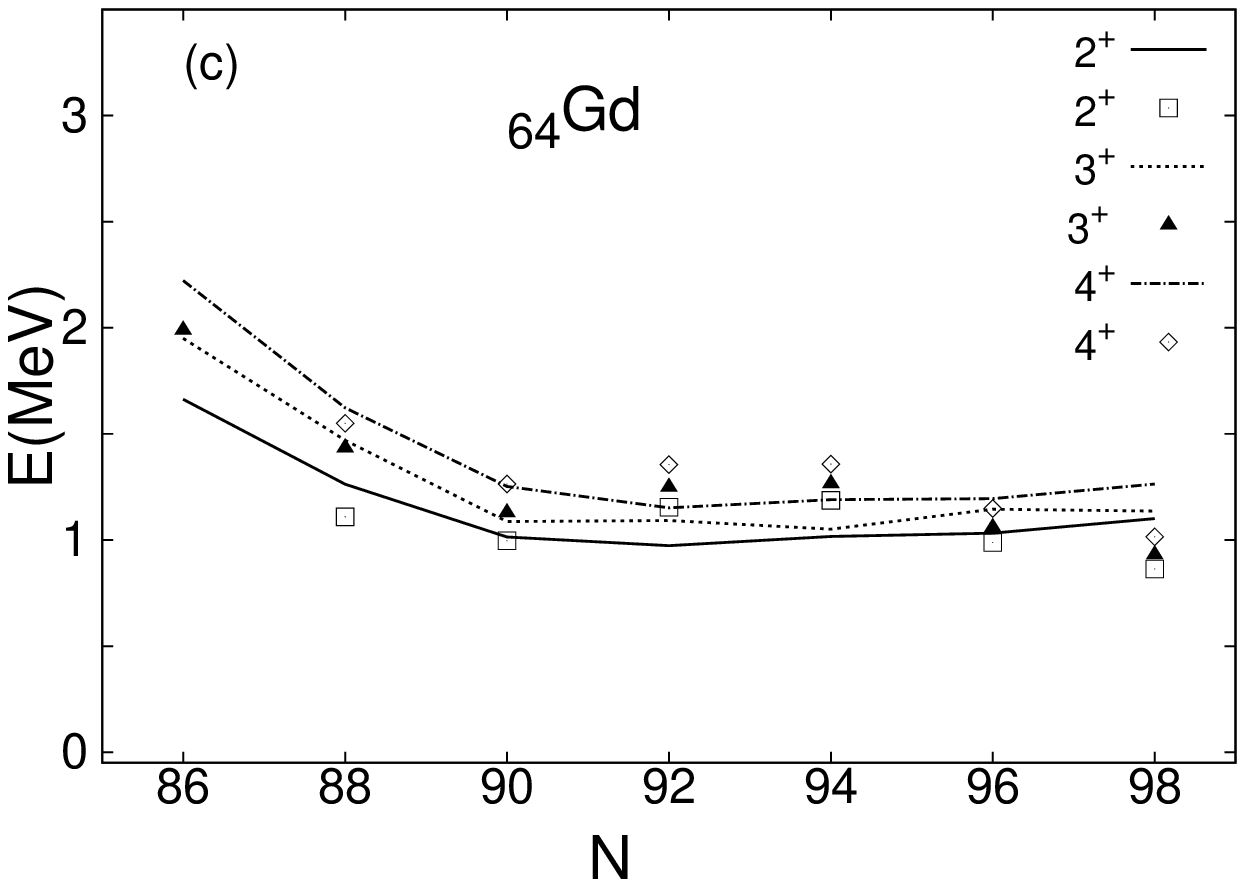}
\includegraphics[width=5.8cm]{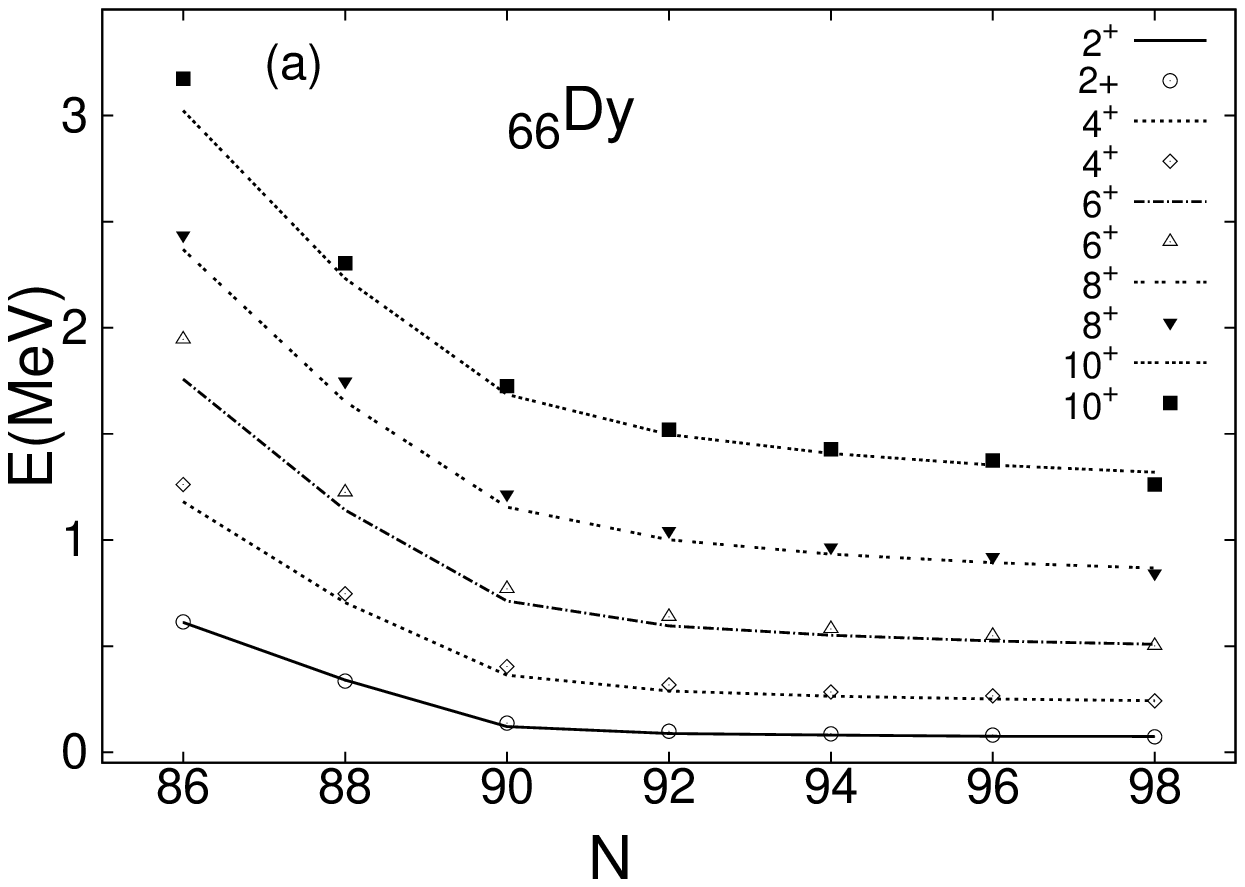}
\includegraphics[width=5.8cm]{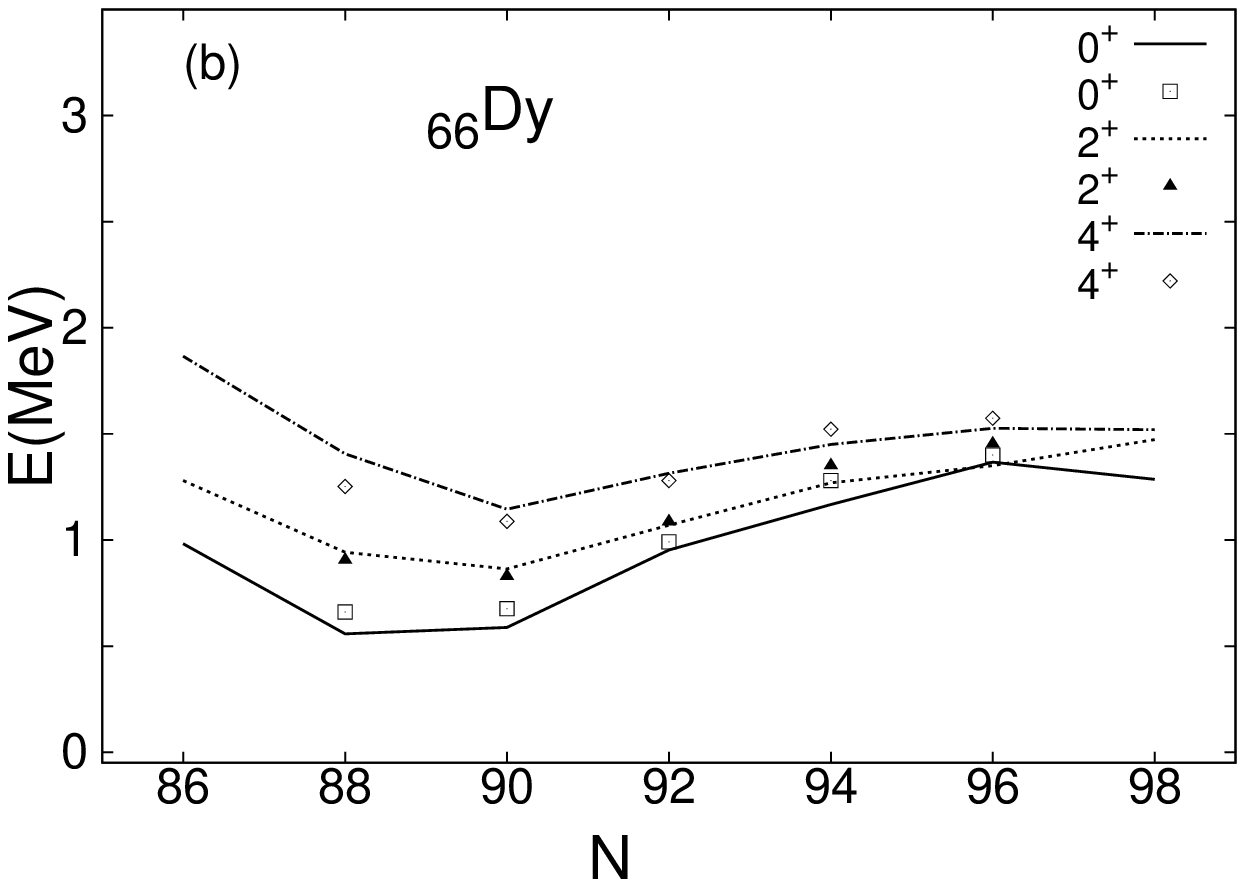}
\includegraphics[width=5.8cm]{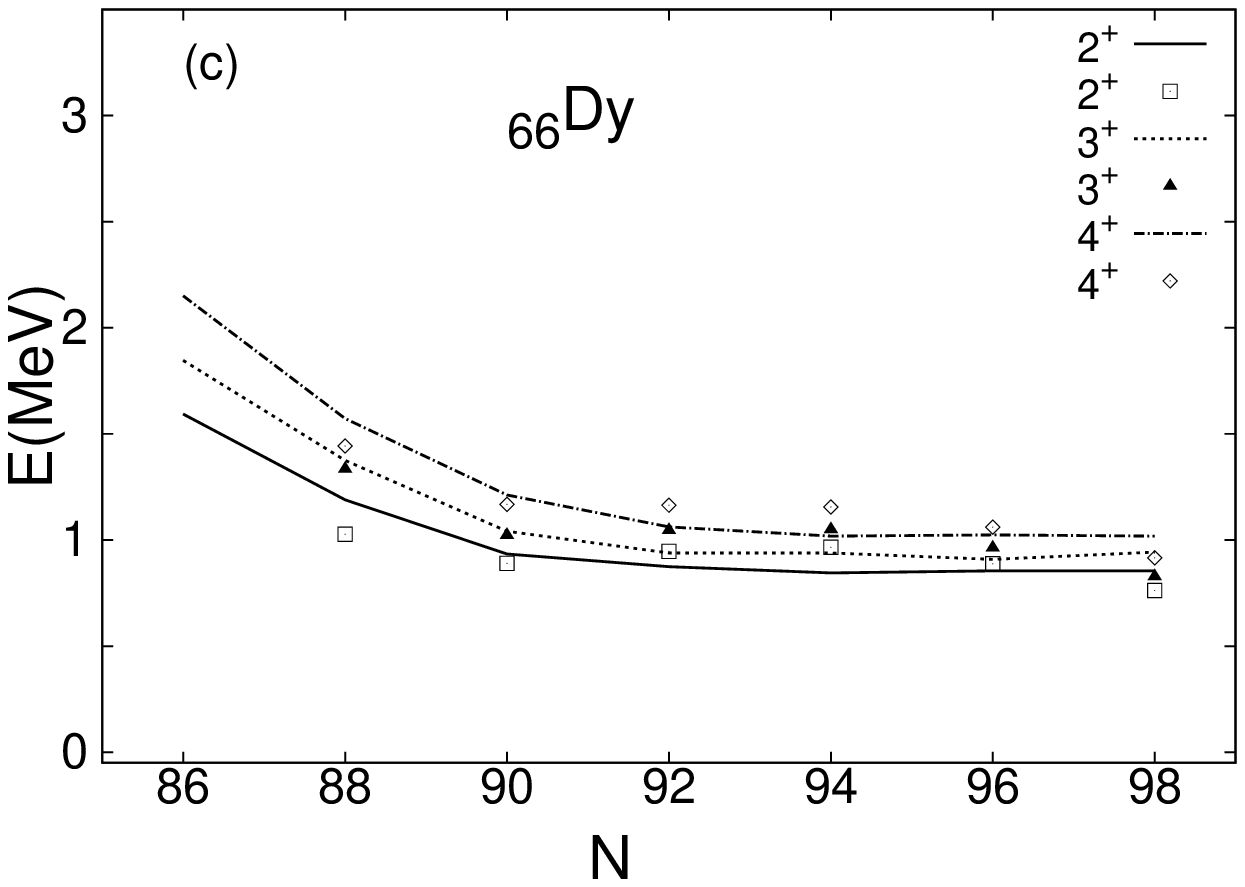}
\caption{
Experimental (points) and calculated (lines) energy levels in the Sm, Gd, and Dy isotopes:
(a) ground-state band up to $J^\pi=10^+$,
(b) first-excited $K^\pi=0^+$ band up to $J^\pi=4^+$,
and (c) first-excited $K^\pi=2^+$ band up to $J^\pi=4^+$.}
\label{f_en}
\end{figure*}
Figure~\ref{f_en} illustrates,
with the examples of samarium, gadolinium, and dysprosium,
the typical evolution of the energy spectrum
of a spherical to that of a deformed nucleus
which is observed for every isotope series studied here.
In each nucleus are shown levels of the ground-state band
up to angular momentum $J^\pi=10^+$
as well as the first few states of the excited bands,
together with their experimental counterparts, if known.

For a given nucleus the choice of the subset of collective states
that should be included in the fit is often far from obvious.
While members of the ground-state and the $\gamma$-vibrational bands
in a deformed nucleus are readily identified,
this is not necessarily so for the `$\beta$-vibrational' band.
As a result, guided by the E0 transitions discussed in the next subsection,
the calculated first-excited $K^\pi=0^+$ band is associated in some nuclei
not with the lowest observed $K^\pi=0^+$ band but with a higher-lying one.
This is the case for $^{168-172}$Yb
and also for $^{166}$Er
where the {\em fourth} $J^\pi=0^+$ level at 1934~keV
has been recently identified
as the band head of the $\beta$-vibrational band
on the basis of its large E0 matrix element to the ground state~\cite{Wimmer09}.

Although the agreement with the experiment can be called satisfactory,
it should be noted from Table~\ref{t_parh} that it is obtained
with rather large parameter fluctuations between the different isotopic chains.
This presumably is so because the parameters are highly correlated
and small changes in the fitted data
give rise to large fluctuations in some of the parameters.
The main purpose of this calculation, however,
is not to establish some parameter systematics with the Hamiltonian~(\ref{e_hgen})
but rather to arrive at a reasonably realistic description
of the spherical-to-deformed transition.
This will enable a simultaneous calculation of charge radii and E0 transitions,
as discussed in the next two subsections.

\subsection{Isotope and isomer shifts}
\label{ss_r2}
\begin{table}
\caption{The parameters $|\alpha|$ and $|\alpha'|$
in the charge radius operators~(\ref{e_r2sd}) and~(\ref{e_r2sda}),
in units of fm$^2$,
for the different isotope series.}
\label{t_parr2}
\begin{tabular}{@{}c|rrrrrrrrr}
\hline
\hline
Isotope&Ce&Nd&Sm&Gd&Dy&Er&Yb&Hf&W\\
\hline
$|\alpha|$~&~0.22~&~0.24~&~0.26~&~0.13~&~0.15~&~0.15~&~0.11~&~0.10~&~0.11\\
$|\alpha'|$~&~0.23~&~0.25~&~0.26~&~0.09~&~0.12~&~0.12~&~0.09~&~0.11~&~0.15\\
\hline
\hline
\end{tabular}
\end{table}
Isotope shifts $\Delta\langle r^2\rangle$, according to Eq.~(\ref{e_ips}),
depend on the parameters $|\alpha|$ and $\eta$
in the $\mbox{IBM-1}$ operator~(\ref{e_r2sd}). 
These parameters are expected to vary smoothly with mass number $A$,
according to Eqs.~(\ref{e_esta}) and~(\ref{e_estb2}).
The estimate~(\ref{e_esta}) neglects, however,
the microscopic make-up of the pair of neutrons
which varies considerably from Ce to W.
It is therefore necessary to adjust $|\alpha|$
for each isotope series separately,
and the resulting values are given in the first row of Table~\ref{t_parr2}.
The value of $\eta$, on the other hand, is kept constant for all isotopes,
$\eta=0.50$~fm$^2$.
The parameters thus derived
are broadly consistent with the estimates
obtained in Subsect.~\ref{ss_estimate1}.

Similar remarks hold for the alternative parametrization~(\ref{e_r2sda})
of the charge radius operator.
The values of $|\alpha'|$ are given in the second row of Table~\ref{t_parr2}
while $\eta'=0.05$~fm$^2$.

\begin{figure}
\centering
\includegraphics[width=7cm]{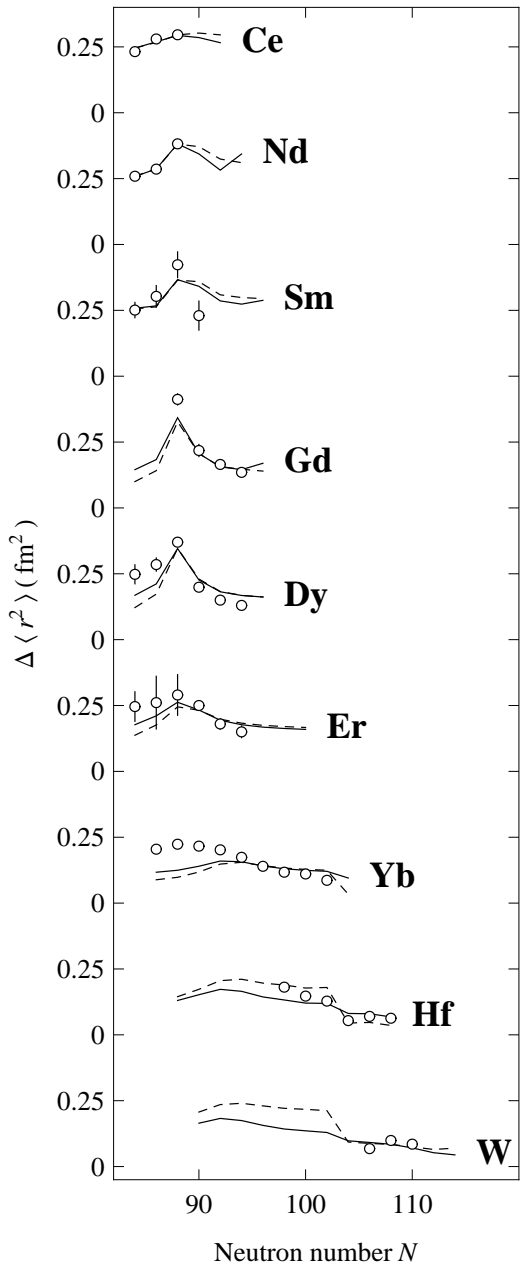}
\caption{
Experimental (points) and calculated (lines) isotope shifts $\Delta\langle r^2\rangle$,
in units of fm$^2$, for isotopic chains
in the rare-earth region from Ce to W.
The full (dashed) lines are for the charge radius operator~(\ref{e_r2sd}) [(\ref{e_r2sda})],
with parameters given in Table~\ref{t_parr2} and in the text.
Data are taken from
Ref.~\cite{Cheal03} for Ce,
from Ref.~\cite{Otten89} for Nd, Sm, Dy, Er, and Yb,
from Ref.~\cite{Frick95} for Gd,
from Ref.~\cite{Angeli04} for Hf,
and from Ref.~\cite{Jin94} for W.}
\label{f_ips} 
\end{figure}
The resulting isotope shifts are shown in Fig.~\ref{f_ips}.
Only mariginal differences exist between the two sets of calculations,
with and without the factor $1/N_{\rm b}$ in the charge radius operator.
The main reason for preferring the form~(\ref{e_r2sd})
is that it lacks a kink in $\Delta\langle r^2\rangle$ at mid-shell.
Although this seems to occur in the Hf data,
it is unlikely that the observed kink is
associated with a maximum of the boson number at mid-shell.
Given the similarity between the two sets of calculations,
the subsequent comments are valid for both.

The peaks in the isotope shifts are well reproduced 
in all isotopic chains with the exception of Yb.
The largest peaks occur for $^{152-150}$Sm, $^{154-152}$Gd, and $^{156-154}$Dy,
that is, for the difference in radii between $N=90$ and $N=88$ isotopes.
The peak is smaller below $Z=62$ for Ce and Nd,
and fades away above $Z=66$ for Er, Yb, Hf, and W.
The calculated isotope shifts broadly agree with these observed features
but there are differences though.
Notably, the calculated peak in the Sm isotopes
is much broader than the observed one,
indicating that the spherical-to-deformed transition
occurs faster in reality than it does in the \mbox{IBM-1} calculation.
Also, it would be of interest to determine
the character of the transition in the Nd isotopes:
the present \mbox{IBM-1} calculation predicts it to be rather smooth
but data in the deformed region of the transition
are lacking to confirm this behavior.
Likewise, the \mbox{IBM-1} calculation features
a very fast transition for the Gd isotopes with a sharp peak at $N=88$
but isotope-shift data are lacking for the nuclei
in the spherical region of the transition.

The feature of peaking isotope shifts is related to the onset of deformation
which is particularly sudden (as a function of neutron number)
for the Sm, Gd, and Dy isotopes.
This can be quantitatively understood
as a result of the subshell closure at $Z=64$~\cite{Ogawa78},
combined with the strongly attractive interaction
between neutrons in the $1\nu h_{9/2}$ orbit
and protons in the $1\pi h_{11/2}$ orbit.
If the occupancies of the neutrons in the $1\nu h_{9/2}$
and of the protons in the $1\pi h_{11/2}$ orbit are both low,
as is the case for $N\leq92$ and $Z\leq64$,
the nucleus is expected to be spherical.
As soon as one of the two orbits becomes significantly occupied,
the strong neutron-proton interaction
will induce occupancy of the partner orbit
and an onset of deformation.

\begin{table}
\caption{Experimental and calculated isomer shifts $\delta\langle r^2\rangle$,
in units of $10^{-3}$ fm$^2$, in the rare-earth region.}
\label{t_ims}
\centering
\begin{tabular}{@{}rc|crcrcrcrlcr}
\hline
\hline
&&&\multicolumn{8}{c}{$\delta \langle r^{2}\rangle$ ($10^{-3}$ fm$^2$)}\\
\cline{4-13}
Isotope&~&~&Th1\footnotemark&~&Th2\footnotemark&~&Th3\footnotemark&~~~&
\multicolumn{2}{c}{Expt}&~&Ref\\
\hline
$^{150}$Sm&&&80&& 72&&    &&49.6\phantom{0}&{\sl2.6}&&\cite{Yamazaki78}\\
$^{152}$Sm&&&37&& 37&&19&&25.\phantom{00}&{\sl7.}&&\cite{Yeboah67}\\
                                 &&&&&&&&&19.\phantom{00}&   &&\cite{Steiner67}\\
                                 &&&&&&&&&14.\phantom{00}&{\sl1.}&&\cite{Bernow68}\\
                                 &&&&&&&&&12.\phantom{00}& &&\cite{Baader68}\\
                                 &&&&&&&&&18.\phantom{00}&{\sl4.}&&\cite{Yamazaki78}\\
$^{154}$Sm&&&  6&&   7&&    &&1.1\phantom{00}&{\sl0.8}&&\cite{Wheeler70}\\
$^{154}$Gd&&& 28&& 31&&&&15.\phantom{00}&{\sl2.}&&\cite{Bernow68}\\
                                 &&&&&&&&&19.\phantom{00}&{\sl6.}&&\cite{Rehm69}\\
                                &&&&&&&&&18.5\phantom{0}&{\sl2.5}&&\cite{Backe74}\\
                                &&&&&&&&&20.0\phantom{0}&{\sl3.9}&&\cite{Laubacher83}\\
$^{156}$Gd&&&   4&&   5&&   && 2.6\phantom{0}&{\sl0.8}&&\cite{Henning68}\\
                                &&&&&&&&& 0.1\phantom{0}&{\sl1.7}&&\cite{Backe74}\\
                                &&&&&&&&& 4.3\phantom{0}&{\sl3.7}&&\cite{Laubacher83}\\
$^{158}$Gd&&&   3&&   4&&   && 0.4\phantom{0}&{\sl0.3}&&\cite{Fink67}\\
                                &&&&&&&&& 1.5\phantom{0}&{\sl0.8}&&\cite{Russell70}\\
                                &&&&&&&&&$-4.0$\phantom{0}&{\sl1.2}&&\cite{Backe74}\\
                                &&&&&&&&& 0.3\phantom{0}&{\sl3.3}&&\cite{Laubacher83}\\
$^{160}$Gd&&&   1&&   1&&   && 0.3\phantom{0}&{\sl0.8}&&\cite{Russell70}\\
                                &&&&&&&&&$-1.4$\phantom{0}&{\sl3.1}&&\cite{Backe74}\\
                                &&&&&&&&&$-2.1$\phantom{0}&{\sl3.2}&&\cite{Laubacher83}\\
$^{170}$Yb&&&   4&&   6&&    &&1.2\phantom{0}&{\sl0.3}&&\cite{Henning68}\\
                                &&&&&&&&&1.7\phantom{0}&{\sl0.6}&&\cite{Henning71}\\
                                &&&&&&&&&1.20&{\sl0.40}&&\cite{Russell73}\\
$^{172}$Yb&&&   3&&   5&&    &&0.41&{\sl0.20}&&\cite{Russell73}\\
$^{174}$Yb&&&   2&&   4&&    &&1.0\phantom{0}&{\sl0.45}&&\cite{Henning71}\\
                                &&&&&&&&&$-0.44$&{\sl0.19}&&\cite{Russell73}\\
$^{176}$Yb&&&   2&&   4&&    &&$-0.17$&{\sl0.10}&&\cite{Russell73}\\
$^{182}$W&&&    5&&   6&&    &&6.0\phantom{0}& &&\cite{Cohen66}\\
                                 &&&&&&&&&$-0.6$\phantom{0}& &&\cite{Baader68}\\
                                 &&&&&&&&&$-0.2$\phantom{0}& &&\cite{Wagner71}\\
$^{184}$W&&&    8&&   9&&    &&0.16& &&\cite{Wagner71}\\
                               &&&&&&&&&0.5\phantom{0}&{\sl0.3}&&\cite{Hardy71}\\
$^{186}$W&&&  17&& 18&&    &&0.14& &&\cite{Wagner71}\\
\hline
\hline
\end{tabular}
\footnotetext[1]{With the charge radius operator~(\ref{e_r2sd}).}
\footnotetext[2]{With the charge radius operator~(\ref{e_r2sda}).}
\footnotetext[3]{Configuration-mixing calculation of Ref.~\cite{Kulp08}.}
\end{table}
A further test of the calculated charge radii
is obtained from isomer shifts $\delta\langle r^2\rangle$,
depending only on $\eta$ [see Eq.~(\ref{e_ims})] or $\eta'$.
The isomer shifts that are known experimentally are listed in Table~\ref{t_ims}.
The data are more than 30 years old and often discrepant.
Nevertheless, a clear conclusion can be drawn
from the isomer shifts measured in the Sm and Gd isotopes:
they are easily an order of magnitude smaller in the deformed
than they are in the spherical region.
In spite of the extreme sensitivity of this effect,
a quantitative description is obtained of the isomer shifts in the Gd isotopes.
For the Sm isotopes only a qualitative agreement is found
since the experimentally observed drop in isomer shift
between $^{152}$Sm and $^{154}$Sm is stronger
than what is calculated in the \mbox{IBM-1}.
This indicates that the spherical-to-deformed transition
is faster in reality than it is  in the calculation,
in line with what can be concluded from the isotope shifts.

From the preceding analysis the following picture emerges.
All considered isotopic chains exhibit an evolution
from a spherical to a deformed shape
which, at the phase-transitional point, 
is characterized by a peak in the isotope shifts.
The height of the peak is proportional to the suddenness of the transition.
This effect is a direct consequence of the increase
in the mean-square radius of a nucleus due to its deformation.
The \mbox{IBM-1} is able
to provide an adequate description of this transitional behavior.
By adjusting the charge radius operator of the \mbox{IBM-1}
to the observed height of the peak in the isotope shifts,
a first estimate of the parameter $\eta$ (or $\eta'$) is obtained.
Its value follows more directly from isomer shifts
since only one parameter enters this quantity
but, unfortunately, data are scarce and often unreliable.
The choice $\eta=0.50$~fm$^2$ (or $\eta'=0.05$~fm$^2$)
is a compromise between the value
obtained from a fit to $\Delta\langle r^2\rangle$ of all isotopes
and the one from $\delta\langle r^2\rangle$ in the Gd isotopes.
The question is now whether this value of $\eta$ (or $\eta'$)
reproduces the E0 transitions observed in the rare-earth nuclei.

\subsection{Electric monopole transitions}
\label{ss_e0}
The calculation of the matrix elements
of the E0 transition operator~(\ref{e_e0sd}) or~(\ref{e_e0sda})
requires the knowledge of the effective charges $e_{\rm n}$ and $e_{\rm p}$.
In principle, an estimate of the ratio $e_{\rm n}/e_{\rm p}$
can be obtained by fitting the expression~(\ref{e_estr2})
to the available data on charge radii in the rare-earth region.
The minimum in the rms deviation is shallow though
and, furthermore, the correlation between $r_0$ and $e_{\rm n}/e_{\rm p}$ is strong.
In other words, a slightly different choice of $r_0$
gives an almost equally good fit to the charge radii of $58\leq Z\leq74$ nuclei
but with a significantly different ratio $e_{\rm n}/e_{\rm p}$.
A reasonable choice of parameters, close to the optimum set,
corresponds to $r_0=1.24$~fm, $e_{\rm n}=0.50e$, and $e_{\rm p}=e$.

\begin{table}
\caption{Experimental and calculated $\rho^2({\rm E0})$ values in the rare-earth region.}
\label{t_e0}
\centering
\begin{tabular}{@{}rc|crcrc|ccc|crcrcrcrl}
\hline
\hline
&&&&&&&&&&&\multicolumn{8}{c}{$\rho^2({\rm E0})\times10^3$}\\
\cline{12-19}
Isotope&~&~&
\multicolumn{3}{c}{Transition}&~&&$J$&~&~~&
Th1\footnotemark&&Th2\footnotemark&&Th3\footnotemark&&\multicolumn{2}{c}{Expt\footnotemark}\\
\hline
$^{150}$Sm&&& 740&$\rightarrow$&    0&&& 0&&&   7&&   6&&     &&  18&{\sl 2}\\
                   &&&1046&$\rightarrow$&334&&& 2&&& 16&& 13&&     &&100&{\sl 40}\\
$^{152}$Sm&&& 685&$\rightarrow$&     0&&& 0&&& 52&& 52&&72&&   51&{\sl 5}\\
                   &&&  811&$\rightarrow$& 122&&& 2&&& 41&& 41&&77&&   69&{\sl 6}\\
                   &&&1023&$\rightarrow$& 366&&& 4&&& 29&& 29&&84&&   88&{\sl 14}\\
                   &&&1083&$\rightarrow$&     0&&& 0&&&   2&&   2&&    &&  0.7&{\sl 0.4}\\
                   &&&1083&$\rightarrow$& 685&&& 0&&& 47&& 47&&    &&  22&{\sl 9}\\
$^{154}$Sm&&&1099&$\rightarrow$&    0&&& 0&&&  41&& 49&&    &&  96&{\sl 42}\\
$^{152}$Gd&&&   615&$\rightarrow$&    0&&& 0&&& 68&& 68&&     &&  63&{\sl 14}\\
                   &&&  931&$\rightarrow$& 344&&& 2&&& 77&& 77&&     &&  35&{\sl 3}\\
$^{154}$Gd&&&  681&$\rightarrow$&     0&&& 0&&& 84&&102&&    &&  89&{\sl 17}\\
                   &&&  815&$\rightarrow$& 123&&& 2&&& 66&& 80&&     &&  74&{\sl 9}\\
                   &&&1061&$\rightarrow$& 361&&& 4&&& 38&& 46&&     &&  70&{\sl 7}\\
$^{156}$Gd&&&1049&$\rightarrow$&     0&&& 0&&& 44&& 64&&      && 42&{\sl 20}\\
                   &&&1129&$\rightarrow$&   89&&& 2&&& 41&& 59&&      && 55&{\sl 5}\\
$^{158}$Gd&&&1452&$\rightarrow$&     0&&& 0&&& 30&& 51&&      && 35&{\sl 12}\\
                   &&&1517&$\rightarrow$&   79&&& 2&&& 27&& 45&&      && 17&{\sl 3}\\
$^{158}$Dy&&&1086&$\rightarrow$&    99&&& 2&&& 42&& 70&&      && 27&{\sl 12}\\
$^{160}$Dy&&&1350&$\rightarrow$&    87&&& 2&&& 28&& 56&&      && 17&{\sl 4}\\
$^{162}$Er&&&1171&$\rightarrow$&   102&&& 2&&& 38&& 64&&      &&630&{\sl 460}\\
$^{164}$Er&&&1484&$\rightarrow$&     91&&& 2&&& 24&& 48&&      && 90&{\sl 50}\\
$^{166}$Er&&&1460&$\rightarrow$&       0&&& 0&&&   9&& 20&&      &&127&{\sl 60}\\
$^{170}$Yb&&&1229&$\rightarrow$&      0&&& 0&&& 32&& 72&&      && 27&{\sl 5}\\
$^{172}$Yb&&&1405&$\rightarrow$&      0&&& 0&&& 30&& 76&&      &&0.2&{\sl 0.03}\\
$^{174}$Hf&&&  900&$\rightarrow$&    91&&& 2&&& 32&& 71&&       && 27&{\sl 13}\\
$^{176}$Hf&&&1227&$\rightarrow$&    89&&& 2&&& 15&& 38&&       && 52&{\sl 9}\\
$^{178}$Hf&&&1496&$\rightarrow$&    93&&& 2&&& 32&& 72&&       && 14&{\sl 3}\\
$^{182}$W&&&1257&$\rightarrow$&  100&&& 2&&& 45&& 77&&       &&3.5&{\sl 0.3}\\
$^{184}$W&&&1121&$\rightarrow$&   111&&& 2&&& 52&& 75&&       &&2.6&{\sl 0.5}\\
\hline
\hline
\end{tabular}
\footnotetext[1]{With the E0 transition operator~(\ref{e_e0sd}).}
\footnotetext[2]{With the E0 transition operator~(\ref{e_e0sda}).}
\footnotetext[3]{Configuration-mixing calculation of Ref.~\cite{Kulp08}.}
\footnotetext[4]{From Ref.~\cite{Kibedi05} for $J=0$,
except $^{154}$Sm and $^{166}$Er which are from Ref.~\cite{Wimmer09};
from Ref.~\cite{Wood99} for $J\neq0$.}
\end{table}
In Table~\ref{t_e0} the available E0 data in the rare-earth region
are compared with the results of this calculation.
The two choices of E0 transition operator, Eqs.~(\ref{e_e0sd}) and~(\ref{e_e0sda}),
again yield comparable results.
An overall comment is that
the present approach succeeds in reproducing
the correct order of magnitude for $\rho^2({\rm E}0)$,
in particular in the Sm, Gd, and Dy isotopes. 
However, some discrepancies can be observed in heavier nuclei
and especially concern $^{172}$Yb and $^{182-184}$W.
A possible explanation is
that the $\rho^2({\rm E}0)$ measured for these nuclei
is not associated with collective states.
This seems to be the case in $^{172}$Yb
where several $\rho^2({\rm E}0)$ have been measured
none of which is large.
Only in the W isotopes does it seem certain that the observed E0 strength
is consistently an order of magnitude smaller than the calculated value.
It is known that these nuclei
are in a region of hexadecapole deformation~\cite{Lee74}
and this may offer a qualitative explanation of the suppression of the E0 strength,
as argued in the next section.

While in a spherical vibrator there is no appreciable E0 strength
from the ground state to any excited $0^+$ state,
this is different in a deformed nucleus
which should exhibit large $\rho^2({\rm E0})$s
from the ground-state towards the $\beta$-vibrational band~\cite{Bohr75,Reiner61}. 
As a consequence, one predicts an increase in the E0 strength
as the phase-transitional point is crossed.
This seems to be confirmed in the few isotopic chains where data are available. 
Adopting a simple, schematic Hamiltonian,
von Brentano {\it et al.}~\cite{Brentano04} showed
that also in the \mbox{IBM-1} sizable E0 strength
should be observed in all deformed nuclei.
The present \mbox{IBM-1} calculation is in qualitative agreement
with this geometric picture and with the results of von Brentano {\it et al.}
Nevertheless, it should be pointed out
that, systematically, the calculated $\rho^2({\rm E0};0^+_2\rightarrow0^+_1)$ in Table~\ref{t_e0}
{\em diminishes} once the phase-transitional point is crossed.
In the Gd nuclei, at least, this behavior seems to be borne out by the data.
It indicates that the first-excited $0^+$ state in the \mbox{IBM-1}
has not simple a $\beta$-vibrational character
but has a more complicated structure~\cite{Garrett01,Casten88}.

Tables~\ref{t_ims} and~\ref{t_e0}
also show the results of a configuration-mixing calculation for $^{152}$Sm~\cite{Kulp08}.
This approach leads to a quantitative, detailed description of the data.
Results of similar good quality are obtained for $^{154}$Gd~\cite{Kulpun}.
However, in view of the employed methodology,
as explained in Subsect.~\ref{ss_coexistence},
it seems difficult to make systematic calculations
of E0 properties of nuclei with this model.

\section{Effect of $g$ bosons on electric monopole transitions}
\label{s_g}
An obvious extension of the $sd$-IBM
is to include a correlated pair of higher angular momentum
for which the most natural choice is the $g$ boson with $\ell=4$.
Many articles have been published over the years
where the role of the $g$ boson has been investigated in detail,
for which we refer the reader to the review of Devi and Kota~\cite{Devi92}.
The $sdg$-IBM has been used
in the interpretation of structural properties of nuclei in the rare-earth region.
For example, properties of the $^{154-160}$Gd isotopes,
including energy spectra and E2, E4, and E0 transitions
were interpreted in the framework of the \mbox{$sdg$-IBM-1}~\cite{Isacker82}.
The conclusion of this particular study,
namely that the $g$ boson is indispensable
for the explanation of the character of some $K^\pi =4^+$ bands,
was confirmed in transfer-reaction studies,
see, {\it e.g.}, Burke {\it et al.}~\cite{Burke01,Burke02}.
Other examples of studies of rare-earth nuclei in the \mbox{$sdg$-IBM-1}
include $^{168}$Er~\cite{Yoshinaga88}, $^{146-158}$Sm~\cite{Devi92b},
and $^{144-150}$Nd~\cite{Perrino93,Devi93}.

In this section the possible influence of hexadecapole deformation
or, equivalently, of $g$ bosons on E0 transitions is discussed.
The spherical-to-deformed shape phase transition in the \mbox{$sdg$-IBM-1}
corresponds to a transition
between the two limits ${\rm U}(5)\otimes {\rm U}(9)$ and SU(3)~\cite{Devi90}.
The following schematic Hamiltonian is adopted:
\begin{equation}
\hat H=\epsilon_d\hat n_d+\epsilon_g\hat n_g-\kappa\hat Q\cdot\hat Q,
\label{e_sdgh}
\end{equation}
where $\hat Q_\mu$ is the ${\rm SU}_{sdg}(3)$ quadrupole operator~\cite{Kota87}
\begin{eqnarray}
\hat Q_\mu&=&
[s^\dag\times\tilde d+d^\dag\times\tilde s]^{(2)}_\mu-
\frac{11}{14}\sqrt{\frac{5}{2}}[d^\dag\times\tilde d]^{(2)}_\mu
\label{e_qsu3}\\&&+
\frac{9}{7}[d^\dag\times\tilde g+g^\dag\times\tilde d]^{(2)}_\mu-
\frac{3}{14}\sqrt{55}[g^\dag\times\tilde g]^{(2)}_\mu.
\nonumber
\end{eqnarray}
For a convenient description of the phase transition,
another para\-metrization of the Hamiltonian~(\ref{e_sdgh}) can be introduced
in terms of $\lambda$ and $\zeta$ (sometimes referred to as control parameters)
which are related to $\epsilon_d$, $\epsilon_g$, and $\kappa$ by
\begin{equation}
\lambda=\frac{\epsilon_g}{\epsilon_d},
\qquad
\kappa=\frac{\zeta}{4N_{\rm b}(1-\zeta)},
\label{e_par}
\end{equation}
where $N_{\rm b}$ is now the total number of $s$, $d$, and $g$ bosons.
The Hamiltonian~(\ref{e_sdgh}) then becomes
\begin{equation}
\hat H=c\left[(1-\zeta)(\hat n_d+\lambda\hat n_g)-\frac{\zeta}{4N_{\rm b}}\hat Q\cdot\hat Q\right],
\label{e_schemh}
\end{equation}
where $c$ is a scaling factor.
The ${\rm U}(5)\otimes {\rm U}(9)$ limit is obtained for $\zeta=0$
whereas the SU(3) limit corresponds to $\zeta=1$.
By varying $\zeta$ from 0 to 1 one will cross the critical point $\zeta_{\rm c}\approx0.5$
at which the spherical-to-deformed transition occurs.

In the \mbox{$sdg$-IBM-1} the charge radius operator is
\begin{equation}
\hat T(r^2)=
\langle r^2\rangle_{\rm c}+
\alpha\hat N_{\rm b}+
\eta\frac{\hat n_d}{N_{\rm b}}+
\gamma\frac{\hat n_g}{N_{\rm b}},
\label{e_r2sdg}
\end{equation}
while the E0 transition operator is
\begin{equation}
\hat T({\rm E0})=
(e_{\rm n}N+e_{\rm p}Z)\left(
\eta\frac{\hat n_d}{N_{\rm b}}+
\gamma\frac{\hat n_g}{N_{\rm b}}\right),
\label{e_e0sdg}
\end{equation}
which are straightforward extensions of the expressions~(\ref{e_r2sd}) and~(\ref{e_e0sd}).
Again, the total number operator $\hat N_{\rm b}=\hat n_s+\hat n_d+\hat n_g$
does not contribute to the E0 transition
and is not included in the operator~(\ref{e_e0sdg}).

Analytic expressions can be derived for the matrix elements
of the operators $\hat n_s$, $\hat n_d$, and $\hat n_g$
for the limiting values of $\zeta$ in the Hamiltonian~(\ref{e_schemh}).
They are known for arbitrary angular momentum $J$
but for simplicity's sake results are quoted for $J=0$ only.
In the ${\rm U}(5)\otimes {\rm U}(9)$ limit they are trivial,
\begin{eqnarray}
\langle0^+_1|\hat n_s|0^+_i\rangle&=&N_{\rm b}\delta_{i1},
\nonumber\\
\langle0^+_1|\hat n_d|0^+_i\rangle&=&0,
\nonumber\\
\langle0^+_1|\hat n_g|0^+_i\rangle&=&0.
\label{e_u59}
\end{eqnarray}
In the SU(3) limit one finds for the ground-state expectation values,
\begin{eqnarray}
\langle0^+_1|\hat n_s|0^+_1\rangle&=&
\frac{N_{\rm b}(4N_{\rm b}+1)}{5(4N_{\rm b}-3)},
\nonumber\\
\langle0^+_1|\hat n_d|0^+_1\rangle&=&
\frac{16(N_{\rm b}-1)N_{\rm b}(4N_{\rm b}+1)}{7(4N_{\rm b}-3)(4N_{\rm b}-1)},
\nonumber\\
\langle0^+_1|\hat n_g|0^+_1\rangle&=&
\frac{64(N_{\rm b}-1)N_{\rm b}(2N_{\rm b}-3)}{35(4N_{\rm b}-3)(4N_{\rm b}-1)},
\label{e_su3r2}
\end{eqnarray}
and for the transition matrix elements
from the ground to the first-excited $0^+$ state,
\begin{widetext}
\begin{eqnarray}
\langle0^+_1|\hat n_s|0^+_2\rangle&=&
\frac{4}{5}
\left[\frac{2(N_{\rm b}-1)N_{\rm b}(2N_{\rm b}-1)(4N_{\rm b}+1)}
{3(4N_{\rm b}-5)^2(4N_{\rm b}-3)}\right]^{1/2},
\nonumber\\
\langle0^+_1|\hat n_d|0^+_2\rangle
&=&\frac{4}{7}
\left[\frac{2(N_{\rm b}-1)N_{\rm b}(2N_{\rm b}-1)(4N_{\rm b}-13)^2(4N_{\rm b}+1)}
{3(4N_{\rm b}-5)^2(4N_{\rm b}-3)(4N_{\rm b}-1)^2}\right]^{1/2},
\nonumber\\
\langle0^+_1|\hat n_g|0^+_2\rangle&=&
-\frac{96}{35}
\left[\frac{2(N_{\rm b}-1)N_{\rm b}(2N_{\rm b}-3)^2(2N_{\rm b}-1)(4N_{\rm b}+1)}
{3(4N_{\rm b}-5)^2(4N_{\rm b}-3)(4N_{\rm b}-1)^2}\right]^{1/2}.
\label{e_su3e0}
\end{eqnarray}
\end{widetext}

It is instructive to compare these results
to the corresponding ones in the \mbox{$sd$-IBM-1}
which is done in Table~\ref{t_clme} in the classical limit $N_{\rm b}\rightarrow\infty$.
\begin{table}
\caption{Matrix elements in the classical limit of SU(3) in the \mbox{$sd$-IBM-1}
and the \mbox{$sdg$-IBM-1}.}
\label{t_clme}
\begin{ruledtabular}
\begin{tabular}{c|ccccccc|cccccc}
&&\multicolumn{5}{c}{$\langle0^+_1|\hat n_\ell|0^+_1\rangle$}
&&&\multicolumn{5}{c}{$\langle0^+_1|\hat n_\ell|0^+_2\rangle$}\\
\cline{3-7}\cline{10-14}
IBM&&$\ell=0$&&$\ell=2$&&$\ell=4$&&&$\ell=0$&&$\ell=2$&&$\ell=4$\\
\hline
$sd$&&
$\displaystyle{\frac{1}{3}N_{\rm b}}$&&
$\displaystyle{\frac{2}{3}N_{\rm b}}$&&---&&&
$\displaystyle{\frac{2}{3}\sqrt{\frac{N_{\rm b}}{2}}}$&&
$\displaystyle{-\frac{2}{3}\sqrt{\frac{N_{\rm b}}{2}}}$&&---\\[2ex]
$sdg$&&
$\displaystyle{\frac{1}{5}N_{\rm b}}$&&
$\displaystyle{\frac{4}{7}N_{\rm b}}$&&
$\displaystyle{\frac{8}{35}N_{\rm b}}$&&&
$\displaystyle{\frac{2}{5}\sqrt{\frac{N_{\rm b}}{3}}}$&&
$\displaystyle{\frac{2}{7}\sqrt{\frac{N_{\rm b}}{3}}}$&&
$\displaystyle{-\frac{24}{35}\sqrt{\frac{N_{\rm b}}{3}}}$\\
\end{tabular}
\end{ruledtabular}
\label{climit_tab}
\end{table}
Considering first the expectation values of $\hat n_\ell$ in the ground state,
one notes that $d$ bosons are dominant in $0^+_1$
both in the $sd$- and \mbox{$sdg$-IBM-1}
and that the contribution of $g$ bosons in the \mbox{$sdg$-IBM-1} is fairly modest.
One therefore does not expect a significant effect
of the $g$ boson on the nuclear radius,
and this should be even more so away from the SU(3) limit
for a realistic choice of boson energies, $0<\epsilon_d<\epsilon_g$.
In other words, the schematic Hamiltonian~(\ref{e_sdgh})
captures the obvious feature that effects of deformation on the nuclear radius
are mainly of quadrupole character,
and that hexadecapole deformation plays only a marginal role.
In terms of model calculations it also means
that the parameter $\gamma$ in the operator~(\ref{e_r2sdg})
is ill determined from radii
because the expectation value of $\hat n_g$ in the ground state is small.
Nevertheless, on physical grounds
one expects $\eta$ as well as $\gamma$ to be positive and of the same order
since both the quadrupole and hexadecapole deformation
have the effect of increasing the nuclear radius.

Turning to the $\langle0^+_1|\hat n_\ell|0^+_2\rangle$ matrix elements,
one notes first of all that in the \mbox{$sd$-IBM-1}
only $\ell=2$ contributes to the $0^+_2\rightarrow0^+_1$ E0 transition
with a coefficient $\eta$ fixed from the isotope shifts.
As discussed in the preceding section,
this typically leads to a large $\rho^2({\rm E0})$
from the `$\beta$-vibrational' to the ground band.
The situation is, however, drastically different in the \mbox{$sdg$-IBM-1}.
It is seen from Table~\ref{t_clme}
that the matrix elements of $\hat n_g$ is larger than that of $\hat n_d$
and of different sign.
Therefore, while changes in the nuclear radius due to the $g$ boson
are expected to be negligible,
one cannot rule out its significant impact
on $\rho^2({\rm E0;0^+_2\rightarrow0^+_1})$ in deformed nuclei.

\begin{figure}
\centering
\includegraphics[width=6cm]{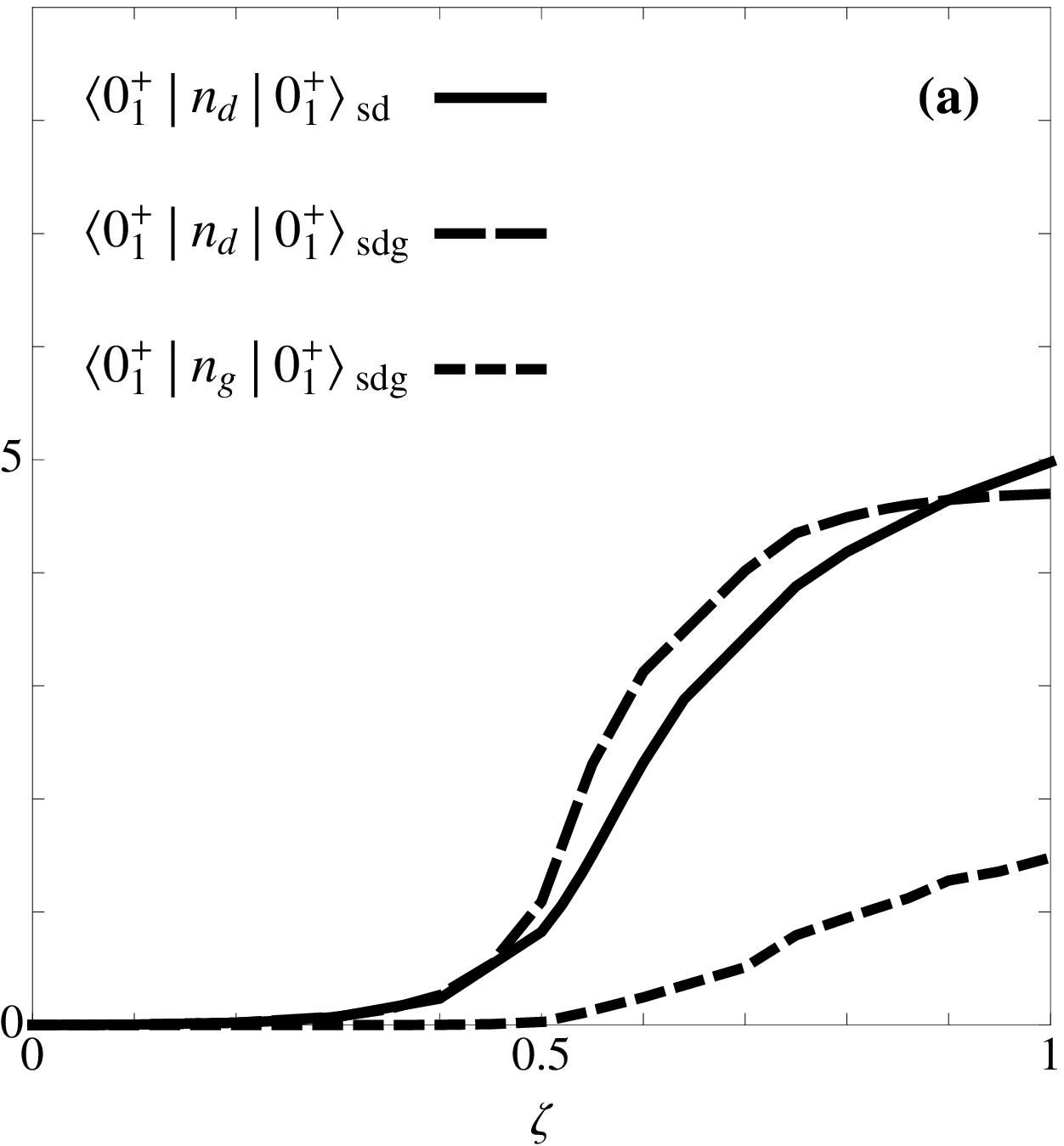}
\includegraphics[width=6cm]{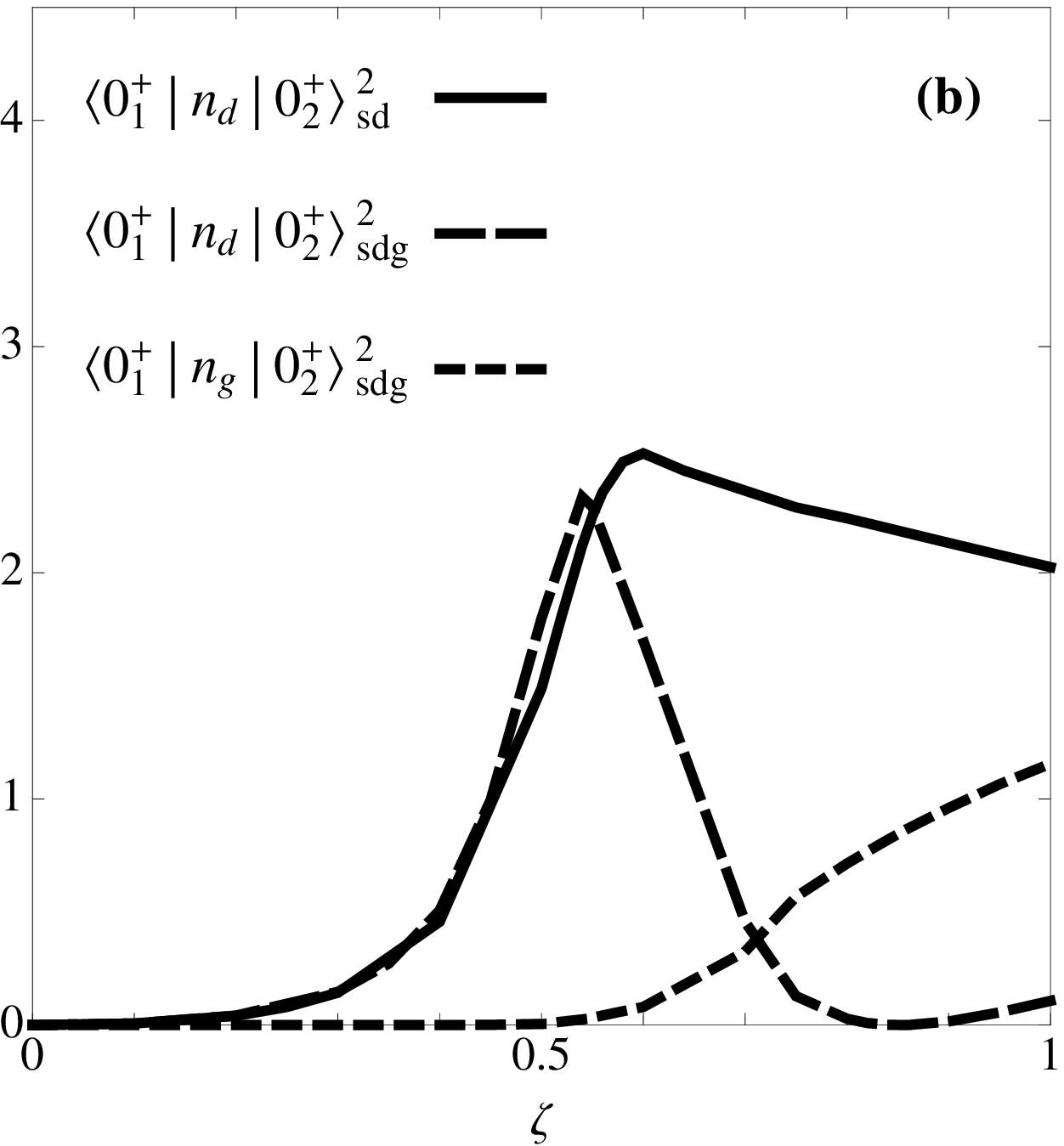}
\caption{
The matrix elements
(a) $\langle0^+_1|\hat n_\ell|0^+_1\rangle$
and (b) $\langle0^+_1|\hat n_\ell|0^+_2\rangle^2$ for $\ell=2$ and $\ell=4$
in the spherical-to-deformed transition of the \mbox{$sd$-IBM-1} and the \mbox{$sdg$-IBM-1}.
In the \mbox{$sd$-IBM-1} the transition is from U(5) to SU(3)
and in the \mbox{$sdg$-IBM-1}
from ${\rm U}(5)\otimes {\rm U}(9)$ to SU(3) with $\lambda\equiv \epsilon_g/\epsilon_d=1.5$.
The number of bosons is $N_{\rm b}=8$.}
\label{f_medg}
\end{figure}
This argument can be made more quantitative
by studying the spherical-to-deformed shape transition
of the Hamiltonian~(\ref{e_sdgh}).
The matrix elements of $\hat n_d$ and $\hat n_g$
can be calculated for arbitrary $\zeta$
with the numerical code {\tt ArbModel}~\cite{Heinzeun}.
A reasonable choice for the ratio of boson energies is $\lambda=1.5$.
The results are shown in Fig.~\ref{f_medg}
and compared to the matrix elements of $\hat n_d$
calculated for the U(5)-to-SU(3) transition in the \mbox{$sd$-IBM-1}.
Figure~\ref{f_medg}(a)
confirms the dominance of the $d$ boson in the ground state of deformed nuclei
both in the $sd$- and \mbox{$sdg$-IBM-1}.
Moreover, the expectation value of $\hat n_d$ varies with $\zeta$
in very much the same way in both models.
In fact, for the entire transition the relation
$\langle0^+_1|\hat n_s|0^+_1\rangle_{sd}\approx\langle0^+_1|\hat n_s|0^+_1\rangle_{sdg}$
approximately holds,
meaning that by choosing $\eta+\gamma$
in the \mbox{$sdg$-IBM-1} equal to $\eta$ in the \mbox{$sd$-IBM-1}
all results of the preceding section concerning radii are reproduced.

In the \mbox{$sd$-IBM-1} as well as in the \mbox{$sdg$-IBM-1}
a sharp increase in $\langle0^+_1|\hat n_d|0^+_2\rangle^2$
is observed around $\zeta_{\rm c}\approx0.5$,
see Fig.~\ref{f_medg}(b).
Up to that point, $\zeta<0.5$, there is essentially no contribution
to $\rho^2({\rm E0};0^+_2\rightarrow0^+_1)$ from the $g$ boson.
Consequently, all \mbox{$sd$-IBM-1} results
up to the phase-transitional point
are not significantly modified by the $g$ boson.
As can be seen from Fig.~\ref{f_medg}(b),
in the deformed regime this is no longer true since, in the \mbox{$sdg$-IBM-1},
a sharp decrease of $\langle0^+_1|\hat n_d|0^+_2\rangle^2$
occurs at $\zeta\approx0.6$ and, furthermore,
$\langle0^+_1|\hat n_g|0^+_2\rangle^2$ rapidly increases
beyond $\zeta\approx0.5$
and dominates $\langle0^+_1|\hat n_d|0^+_2\rangle^2$ for $\zeta\geq0.7$.
The explanation of this dominance is
that $\langle0^+_1|\hat n_d|0^+_2\rangle$ changes sign
before reaching its value in the ${\rm SU}_{sdg}(3)$ limit,
in agreement with the analytical results quoted above.

\section{Conclusions}
\label{s_conc}
In this paper we proposed a consistent description
of nuclear charge radii and electric monopole transitions.
The ingredients at the basis of such a description are
(i) the derivation of a relation between the effective operators describing
nuclear charge radii and electric monopole transitions,
(ii) the mapping of these operators
from the shell model to the interacting boson model,
(iii) the description of spectroscopic properties of chains of isotopes
through the shape-transitional point with the interacting boson model,
and (iv) the assumption that initial and final states
in the considered electric monopole transitions have a collective character
and can be adequately described with the interacting boson model.

The validity of this approach was tested
with an application in even-even nuclei in the rare-earth region ($58\leq Z\leq74$)
which systematically display a spherical-to-deformed transition.
This transitional behavior could be successfully reproduced
with the interacting boson model
and was shown to be correlated with peaks in the isotope shifts,
as observed at the phase-transitional point.
In particular, the correlation between the suddenness of the shape transition
and the height of the peak in the isotope shift
could be correctly reproduced by the model.
With the charge radius operator determined in this way from isotope and isomer shifts,
an essentially parameter-free and systematic calculation
of electric monopole transitions in the rare-earth region
could be undertaken.
The observed electric monopole strengths
were reproduced to within a factor 3,
except in the isolated case of $^{172}$Yb and in the W isotopes.
As a possible explanation for the failure of the approach in the latter isotopes,
the role of hexadecapole deformation or, equivalently, of the $g$ boson
was explored in a schematic model.
It was concluded that the effect of the $g$ boson is marginal on charge radii
but can be strong on electric monopole transitions.

We are aware that our explanation of electric monopole strength
is based on a geometric picture of the nucleus,
in contrast to an alternative explanation
in terms of shape coexistence and configuration mixing.
While there are undoubtedly regions of the nuclear chart
({\it e.g.}, Sr, Zr, and Mo isotopes)
where the latter mechanism is needed to explain the observed electric monopole strength,
we have presented here a comprehensive analysis
of this quantity in rare-earth nuclei
that lends support to the geometric interpretation.

Many thanks are due to Stefan Heinze
who helped us with the numerical calculation involving the $g$ boson.
We also thank Ani Aprahamian, Lex Dieperink, Kris Heyde, and John Wood
for stimulating discussions.
This work was carried out in the framework of CNRS/DEF project 19848.
S.Z.\ thanks the Algerian Ministry of High Education and Scientific Research for financial support.
This work was also partially supported by the
Agence Nationale de Recherche, France,
under contract ANR-07-BLAN-0256-03.

\end{document}